\documentclass[review]{elsarticle}

\usepackage{lineno,hyperref}
\usepackage{verbatim}
\usepackage{placeins}
\usepackage{graphicx}
\usepackage{subcaption}
\modulolinenumbers[5]
\usepackage{amsmath}
\journal{Astronomy and Computing}

\biboptions{authoryear}

\usepackage[left=2cm,top=2cm,right=2cm,bottom=2cm,footskip=1cm]{geometry}
\usepackage{multirow}
\usepackage{diagbox}
\begin{document}
\begin{frontmatter}

\title{Habitability Classification of Exoplanets: A Machine Learning Insight}


\author{Suryoday Basak}
\address{University of Texas at Arlington}

 \author{Surbhi Agrawal, Snehanshu Saha, Abhijit Jeremiel Theophilus, Kakoli Bora}
 \address{PES University South Campus}

 \author{Gouri Deshpande}
 \address{University of Calgary}

 \author{Jayant Murthy}
 \address{Indian Institute of Astrophysics}





\begin{abstract}
We explore the efficacy of machine learning (ML) in characterizing exoplanets into different classes. The source of the data used in this work is University of Puerto Rico's Planetary Habitability Laboratory's Exoplanets Catalog (PHL-EC). We perform a detailed analysis of the structure of the data and propose methods that can be used to effectively categorize new exoplanet samples. Our contributions are two fold. We elaborate on the results obtained by using ML algorithms by stating the accuracy of each method used and propose the best paradigm to automate the task of exoplanet classification. The exploration led to the development of new methods fundamental and relevant to the context of the problem and beyond. Data exploration and experimentation methods also result in the development of a general data methodology and a set of best practices which can be used for exploratory data analysis experiments.

\end{abstract}

\begin{keyword}
machine learning, exoplanets, habitability, elastic gini coefficient, thermal suitability score.
%\texttt{elsarticle.cls}\sep \LaTeX\sep Elsevier \sep template
%\MSC[2010] 00-01\sep  99-00
\end{keyword}

\end{frontmatter}

%\linenumbers

\section{Introduction} \label{sec:intro}

For hundreds of years, astronomers and philosophers have considered the possibility that the Earth is a very rare case of a planet as it harbors life. This is partly because so far, missions exploring specific planets like Mars and Venus have found no traces of life. However, over the past two decades, discoveries of exoplanets have poured in by the hundreds and the rate at which exoplanets are being discovered is increasing. The inference from this is that planets around stars are a rule rather than an exception with the actual number of planets exceeding the number of stars in our galaxy by orders of magnitude \citep{Strigari2012}. Scientists are now discussing the conditions, circumstances, and various possibilities that can lead to the emergence of life \citep{Bains2016} order to find interesting samples from the massive ongoing growth in the data, a sophisticated computational pipeline should be developed which can quickly and efficiently classify exoplanets based on habitability classes. 

The process of discovery of exoplanets is rather complex as the size of exoplanets is small compared to other types of stellar objects such as stars, galaxies, quasars, etc. which can be discovered with greater ease. Given the rapid technological improvements and the accumulation of a large amount of data, it is pertinent to explore advanced methods of data analysis to rapidly classify planets into appropriate categories based on the physical characteristics. Existing work on characterizing exoplanets are based on assigning habitability scores to each planet which allows for a quantitative comparison with Earth. The Biological Complexity Index (BCI) \citep{Irwin2014} , the Earth Similarity Index (ESI) and the Planetary Habitability Index (PHI) \citep{SchulzeMakuch2011} are distance-based metrics which gauge the similarity of a planet to that of Earth; the Cobb-Douglas Habitability Score (CDHS) \citep{CDHPF2016} makes use of econometric modeling to quantify the potential of habitability of an exoplanet. Recently, NASA announced the confirmation of two exoplanets using artificial intelligence \citep{googlenasa}. \cite{Proxb} used an advanced tree-based classifier, Gradient Boosted Decision Trees (GBDT) \citep{Friedman00greedyfunction,Chen2016} to classify Proxima b and planets in the TRAPPIST-1 system. The accuracies were nearly perfect, giving us the basis of exploring other machine classifiers and approaches for the task.

In this paper, we explore the efficacy of different ML approaches to classify exoplanets into thermal habitability classes \citep{thermalclassif} and characterize them based on potential habitability. We try different families of classifiers: probabilistic, instance-based, hard-boundary and tree-based. We emphasize on the challenges with the dataset which include the dominance in the dataset of non-habitable planetary samples and we address these challenges using methods of under sampling (of samples from the dominant class) and oversampling (of samples from the underrepresented class). Relating to the structure of the data and the greater variance in the attributes of the non-habitable samples, we have developed a new splitting criteria \citep{Quinlan1986} which can be used in tree-based classifiers to preferentially classify samples of classes of our choice. The criteria that we have developed is called the \textit{elastic Gini} criteria which can be used to determine the best split at a node in a decision tree for classification. To characterize exoplanets, we have developed a novel Thermal Suitability Score (TSS) which provides a number that does the job of discrimination of potentially habitable planets from non-habitable planets. The basis of this metric is the surface temperature; an appropriate feature extraction is done prior to the computation of the score. Overall, the goal is to develop a pipeline based on various approaches which can quickly detect interesting exoplanetary samples in large databases, thus accelerating the process of discovery of exoplanets.

Our intention of comparing the results of various methods and providing explanations for the same is to demonstrate that the method of classification which works the best is very closely related to what we can understand from the data: how the samples of various classes are distributed and how various features are related. To the best of our knowledge, such an exploration has not been done prior to the current work. %The authors of (\textbf{cite Proxima b paper}) have shown the efficacy of GBDT for the purpose of classification of a small sample of exoplanets; here, we extend and elaborate on the approaches used and we affirm the choice of GBDT as the best classifier for the given dataset.\\

\begin{figure}[htbp!]
\begin{center}
\includegraphics[width=0.8\columnwidth]{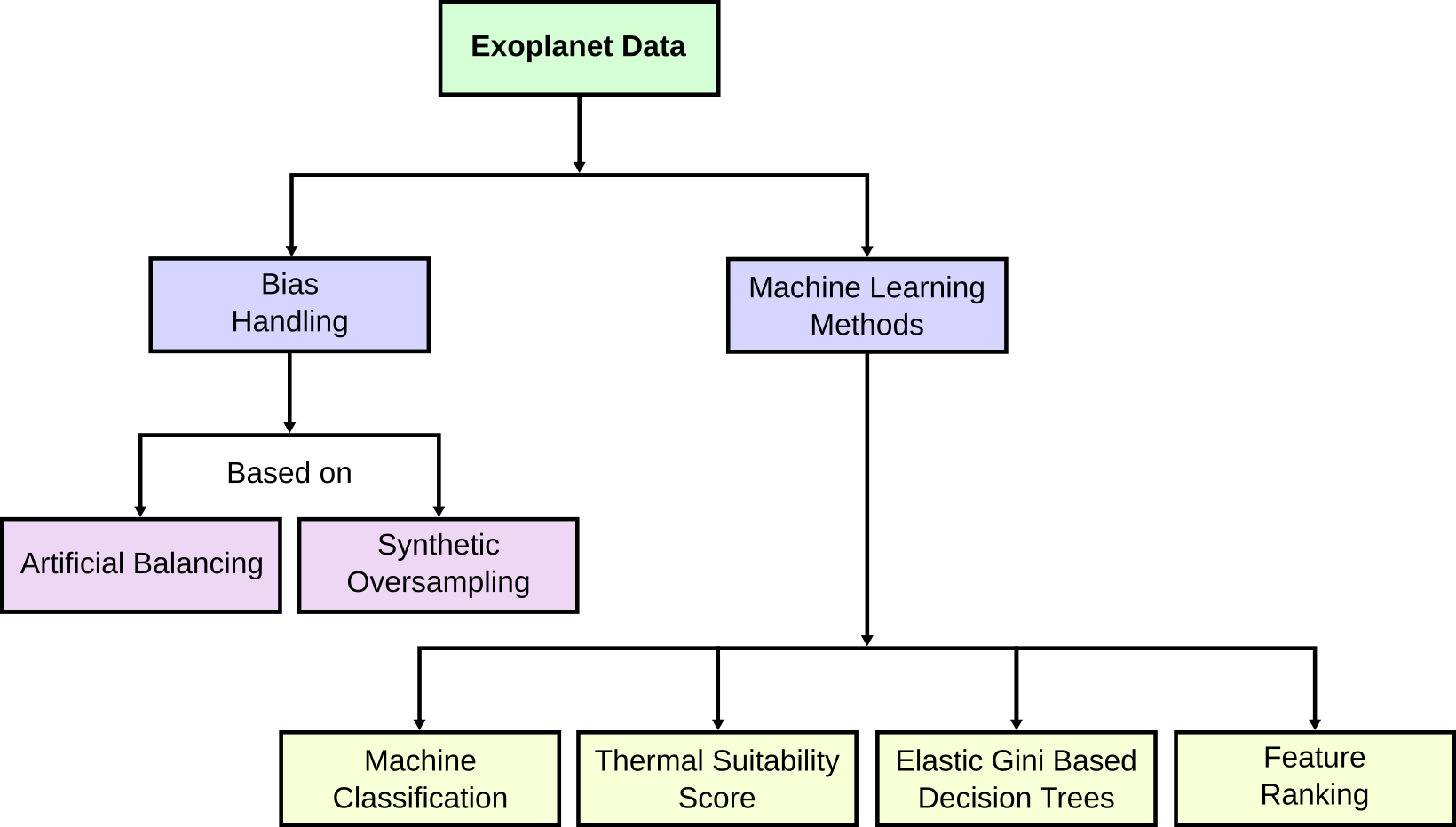}
\caption{An overview of the methods used in the paper.}
\label{fig:overview}
\end{center}
\end{figure}

 %Beyond a comprehensive methodological review, the paper presents techniques and improvisations that are important contributions to the literature.
 
The remainder of the paper is organized as follows. In Section \ref{sec:data}, we briefly present the structure of the PHL-EC data and the complexities associated with a machine learning exploration of this dataset. We build up on the novelty of the presented work in Section \ref{sec:contrib}, where we summarize the contributions of the current work. In Section \ref{sec:methods-existing}, we elaborate on the generic or existing methods in ML that we have used in the current work along with their results. In Section \ref{sec:methods-novel}, we elaborate on the methods we developed while specifically exploring challenges and nuances in the PHL-EC dataset, along with their results. In Section \ref{sec:discussion}, we justify the outcome of the current work and we conclude with some final remarks in Section \ref{sec:conclusion}. Further explanations and elaboration of some of the methods in the current work (and their continuation in the future) can be found in \ref{app:imbalance}, \ref{app:data_aug} and \ref{app:sbaf}.

\section{Data} \label{sec:data}

%PHL-EC has been derived from the Hipparcos catalog by examining the information on distances, stellar variability, multiplicity, kinematics, and spectral classification. 

Combined with stellar data from the Hipparcos catalog \citep{hipparcosref}, the PHL-EC dataset \citep{phlref,SchulzeMakuch2011} consists of a total of 68 features (of which 13 are categorical and 55 are continuous valued) and more than $3800$ confirmed exoplanets (at the time of writing of this paper). The PHL-EC  consists of observed as well as derived attributes. Hence, it presents interesting challenges from the analysis point of view. There are six classes in the dataset, of which we can use three in our analysis as they are sufficiently large in size. These are namely non-habitable, mesoplanet, and psychroplanet classes. These three classes or types of planets can be described on the basis of their thermal properties as follows:

\begin{enumerate}
\item \textbf{Mesoplanets}: The planetary bodies whose sizes lie between Mercury and Ceres falls under this category (smaller than Mercury and larger than Ceres). These are also referred to as M-planets. These planets have mean global surface temperature between 0$^\circ$C to 50$^\circ$C, a necessary condition for complex terrestrial life. These are generally referred as Earth-like planets.
\item \textbf{Psychroplanets}:  These planets have mean global surface temperature between -50$^\circ$C to 0$^\circ$C. Hence, the temperature is colder than optimal for sustenance of terrestrial life.
\item \textbf{Non-Habitable}: Planets other than mesoplanets and psychroplanets do not have thermal properties required to sustain life.
\end{enumerate}

The remaining three classes in the data are those of thermoplanet, hypopsychroplanet and hyperthermoplanet. However, at the time of writing this paper, the number of samples in each of these classes is too small (each class has less than 3 samples) to reliably take them into consideration for an ML-based exploration.

%The catalog includes important features like atmospheric type, mass, radius, surface temperature, escape velocity, earth's similarity index, flux, orbital velocity etc. \textbf{An important aspect is that the values of eccentricity are reported as 0 in cases where the values are actually unknown (and we use values reported exactly as they are present)}. The online data source for the current work is available at \href{http://phl.upr.edu/projects/habitable-exoplanets-catalog/data/database}{http://phl.upr.edu/projects/habitable-exoplanets-catalog/data/database}.

The catalog includes important features like atmospheric type, mass, radius, surface temperature, escape velocity, earth's similarity index, flux, orbital velocity etc. As a first step, data from PHL-EC is  pre-processed. An important challenge in the dataset is that a total of about $1\%$ of the data is missing (with a majority being of the feature P. Max Mass) and in order to overcome this, we used a simple method of substituting the missing data using the class-wise mean of the respective feature, (except for P. Max Mass, which we dropped as a field). Many columns in the data (we shall hereon address these appropriately as \textit{features}) are naturally inter-correlated. Certain attributes from the database namely P.NameKepler (planet's name), Sname HD and Sname Hid (name of parent star), S.constellation (name of constellation), Stype (type of parent star), P.SPH (planet standard primary habitability), P.interior ESI (interior earth similarity index), P.surface ESI (surface earth similarity index), P.disc method (method of discovery of planet), P.disc year (year of discovery of planet), P. Max Mass, P. Min Mass, P.inclination and P.Hab Moon (flag indicating planet's potential as a habitable exomoons) were removed as these attributes do not contribute to the nature of classification of habitability of a planet. Interior ESI and surface ESI, however, together contribute to habitability, but since the dataset directly provides P.ESI, these two features were neglected. Following this, the ML approaches were used on the processed dataset. In all, 45 features were used. The data flow diagram of the entire system is depicted in Figure \ref{fig:overview}. The online data source for the current work is available at \href{http://phl.upr.edu/projects/habitable-exoplanets-catalog/data/database}{http://phl.upr.edu/projects/habitable-exoplanets-catalog/data/database}.

\section{Contributions} \label{sec:contrib}

Our contribution are largely of two kinds:

\begin{enumerate}

\item \textbf{Application of Existing Methods in Machine Learning}: We use existing families of classifiers and data processing methods which are popular in literature. The purpose of this is twofold: first, to explore how the well known and well tested methods can be used to automate the classification of exoplanets and second, to set a performance baseline for new methods that will be subsequently developed for the same tasks. 

%\item \textbf{Novel Classification Methods}:
\item \textbf{Novel Methods of Characterization and Classification}: Based on the nuances in the dataset and the problem statement, and the results of existing methods, we have developed new methods to classify and characterize samples of exoplanets. These are the SVM-KNN based oversampling method, the \textit{elastic} Gini splitting criteria, and the Thermal Suitability Score. To establish efficacy and relevance of the proposed methods, results are compared with, where ever appropriate, the existing methods in literature.

\end{enumerate}
The transition from one item to the other is effortless as demonstrated later in the manuscript. The effort was initially meant at exploring different classification algorithms in the study of automated discrimination of habitability. The exercise evolved into writing novel methods as the authors felt strong reasons to improve upon the existing ones, in the context of the problem. In all the results that we have reported, we have related the performance of the different classifiers to the nature and structure of the data. The algorithms are not treated as `black-boxes' and are thoroughly examined for the purpose of determining the appropriateness of application to the PHL-EC dataset. The working principles of each classifier we have tried are not the same and we have provided justification for the results obtained.

\section{Application of Existing Methods in Machine Learning} \label{sec:methods-existing}

We briefly describe methods popular in literature that we have used.

\subsection{Existing Methods Used} \label{subsec:methods-existing}
\subsubsection{Machine Learning Algorithms} \label{subsubsec:ml-algos}

The following are the ML classification algorithms were used in the current work.

\begin{enumerate}[a.]

\item \textbf{Probabilistic Classifiers}: {\it Gaussian Na{\"i}ve Bayes} evaluates the classification labels based on class-conditional probabilities with class apriori probabilities \citep{rish2001empirical}. GNB works on the assumptions that the features are independent of each other and that they all come from a Gaussian distribution.

\item \textbf{Instance-Based Classifiers}: The {\it k-nearest neighbor} classifier is an instance-based classifier where the distance between the \textit{neighbors} in the input space is used as a measure for categorization \citep{Mohanchandra2015}. Here, $k$ is the number of \textit{nearest} neighbors which are considered for automated classification: the class with the largest number of instances within the nearest $k$ neighbors is predicted as the class of a test sample. We considered $k$ to be 3 while the weights are assigned uniform values.

\item \textbf{Hard-Boundary Classifiers}: The working principle of \textit{support vector machines} is to construct hard-boundaries which are $n$-dimensional hyperplanes, between the samples of different classes \citep{vapnik1964,Cortes1995}. Here, $n$ is the dimensionality of the feature space of the data as fed to the machine. We tried SVM without a kernel, and with a radial basis kernel \citep{Boser1992}.

\item The parameters setup for {\it linear discriminant analysis} classifier was implemented by the decomposition strategy similar to {\it SVM} \citep{duda2001pattern}. No shrinkage metric was specified and no class prior probabilities were assigned. 

\item \textbf{Tree-Based Classifiers}: {\it Decision trees} build tree based data structures by using a criterion, namely \textit{information gain} (or simply, gain)\citep{Quinlan1986}. The discrimination which leads to the best value of gain is used to develop a \textit{rule}; should the rule not result in a perfect discrimination of the data (which is often a consequence of accepting the best rule), then the criterion is recursively applied to the partitions in the data created by the previous rule. A \textit{random forest} is an ensemble of many decision trees \citep{Breiman2001} and \textit{gradient boosted decision trees} are further sophisticated as they do not consider the data in their raw form but use a functional approximation prior to discrimination \citep{ Friedman00greedyfunction}. The splitting criteria that we used were Gini and elastic Gini (discussed in Section \ref{subsec:egini})

%\item \textit{XGBoost} is a recent ensemble tree-based method which optimizes the tree being built. For this algorithm, the maximum number of estimators chosen to develop a classifier was 1000 and the maximum permissible depth of each tree bound at 8. The objective function used was that of a multinomial softmax.

\end{enumerate}

We address the complexity and attribute correlations in the dataset in a two-fold manner: by considering all the features in the dataset in one set of automated classification experiments, and by considering only the basic observables of mass and radius in another set of experiments. The accuracies of either set of experiments are presented and discussed.

The existing methods were implemented using the scikit-learn module in python \citep{scikit-learn}, except GBDTs, which was implemented using XGBoost \citep{Chen2016}.

\subsubsection{Artificial Balancing by Undersampling} \label{subsubsec:undersampling}

In order to mitigate the effects of bias introduced in the data which permeates to the results, we randomly down-sampled the non-habitable class for each test run as it has a dominating number of samples over the other two classes in the data. We call this process \textit{undersampling} as the fundamental idea is to experiment with datasets which do not suffer from biases.

Since the number of psychroplanets and mesoplanets are smaller and comparable in magnitude, we consider all these samples for each run of the experiment, and we randomly sample $5\%$--$10\%$ of the non-habitable samples. The classification methods are then applied to a balanced dataset so constructed. The balancing and classification is repeated many times so that the results finally obtained are representative of the average-case classification. Thus, the results we report in Sections \ref{subsec:results-existing} and \ref{subsec:results-classif-novel} is the average of many classification and testing runs, and each run was done on a dataset with a smaller, random sample of non-habitable planets.

\subsubsection{Synthetic Oversampling Using Parzen Window Estimation} \label{subsubsec:parzen_windows}

Another approach to handle the effects of bias due to the non-habitable class is an artificial data augmentation or \textit{oversampling}. Here, we try to artificially add samples to the classes with lesser number of samples. Essentially, the paradigm of choice is to analyze the class-wise distribution of the data and to generate reasonable samples which reflect the general properties of their respective classes.

%\subsubsection{Estimating a Density Function}

As a part of this approach, we estimate the density of the data by approximating a distribution function empirically -- we do not assume that the numeric values in the data samples are drawn from a standard probability distribution. The method we use for this is known as \textit{window estimation}. This approach was developed by Rosenblatt and Parzen \citep{Rosenblatt1956,Parzen1962}. In a broader sense, window estimation is a method of \textit{kernel density estimation} (KDE).

Here we augment variables from the top 85\% of the features based on their importance as determined by random forest classifiers (Table \ref{tab:feat-imp}). The experiments using KDE is done in the same manner as the experiments using undersampling.

%=======================================================================

\subsection{Results} \label{subsec:results-existing}

The classification algorithms are tested after a method of handling class-imbalance has been used on the data. In order to evaluate the classifiers, we present the confusion matrices for each classifier after preprocessing. The confusion matrices can quantitatively indicate the proportion of instances of each class that are classified correctly or incorrectly. The $(i,j)^{th}$ entry in a confusion matrix indicates the percentage of the times a sample belonging to the $i^{th}$ is classified by an algorithm into the $j^{th}$ class. For a comparison baseline, we also present the results with no undersampling or data augmentation (\ref{app:imbalance}).

\begin{enumerate}

\item \textbf{Artificial Balancing by Undersampling}: After artificially balancing the dataset, the experiments were performed with different feature sets. In the first set of experiments, all the features after the removal of the unimportant features were used. This feature set is expansive and the classification algorithms are expected to find the best patterns from this data. In the second set of experiments, we use only mass and radius as features and this is more reflective of the classification being done without surface temperature.

\item \textbf{Synthetic Oversampling Using Kernel Density Estimation}: KDE was used to generate 1000 psychroplanet and mesoplanet samples. After the data was generated, the classifiers were trained and tested. Here too, the experiments were done separately using the entire feature set, and only mass and radius.

\end{enumerate}

\begin{comment}

\subsubsection{Using An Extensive set of Features}

As a lot of the attributes in the dataset are derived from mass, radius and distance from the parent star \textbf{I think this is right...}, the correlations in the different features could influence the results of classification.

\subsubsection{Using A Restrictive set of Features}

Here, only the minimum mass, mean mass and radius were used as features for classification. This feature set is bereft of the majority of derived planetary parameters and a good performance would indicate the existence of patterns in the data which can be used for sophisticated categorization of exoplanets using just the basic observables.

\end{comment}

The results are presented in Table \ref{table:big}. The results are to be interpreted for each method of oversampling with the feature set used to test the classifiers.

\begin{table*}[!htbp]
\centering
\resizebox{\textwidth}{!}{
\begin{tabular}{| c | c | c c c | c c c | c c c | c c c |}
\hline
\multirow{4}{*}{Algorithm} & & \multicolumn{12}{c|}{Method of Handling Bias due to Class Imbalance} \\ \cline{3-14}
 & & \multicolumn{6}{c|}{Artificial Undersampling} & \multicolumn{6}{c|}{KDE Using Parzen Window Estimation} \\ \cline{2-14}
 & \multirow{2}{*}{\backslashbox{True}{Pred}} & \multicolumn{3}{c|}{All Features} & \multicolumn{3}{c|}{Mass and Radius Only} & \multicolumn{3}{c|}{All Features} & \multicolumn{3}{c|}{Mass and Radius Only} \\ \cline{3-14}

 &  & N & P & M & N & P & M & N & P & M & N & P & M\\
\hline

& N & 99.51 & 0.25 & 0.25 & 96.34 & 1.13 & 2.53 & 99.63 & 0.17 & 0.2 & 96.96 & 1.86 & 1.18 \\
GNB & P & 0.85 & 98.87 & 0.28 & 6.96 & 8.86 & 84.18 & 0.0 & 100.0 & 0.0 & 0.0 & 38.5 & 61.5 \\
 & M & 0.0 & 6.54 & 93.46 & 0.0 & 7.13 & 92.87 & 0.0 & 0.08 & 99.92 & 0.0 & 5.92 & 94.08 \\
\hline
& N & 97.32 & 1.18 & 1.5 & 94.91 & 0.4 & 4.69 & 98.87 & 0.2 & 0.93 & 93.64 & 3.37 & 2.99 \\
LDA & P & 0.31 & 67.18 & 32.51 & 17.8 & 0.0 & 82.2 & 0.0 & 89.06 & 10.94 & 0.0 & 41.51 & 58.49 \\
 & M & 0.0 & 3.93 & 96.07 & 10.94 & 0.0 & 89.06 & 0.0 & 5.62 & 94.38 & 0.0 & 21.96 & 78.04 \\
\hline
& N & 98.12 & 0.52 & 1.36 & 96.39 & 0.94 & 2.67 & 99.65 & 0.12 & 0.23 & 95.97 & 2.36 & 1.68 \\
SVM & P & 3.01 & 86.07 & 10.93 & 6.77 & 20.31 & 72.92 & 0.01 & 99.78 & 0.21 & 0.0 & 70.06 & 29.94 \\
 & M & 3.02 & 2.01 & 94.97 & 0.16 & 13.46 & 86.38 & 0.01 & 0.08 & 99.92 & 0.0 & 29.16 & 70.84 \\
\hline
 & N & 100.0 & 0.0 & 0.0 & 96.87 & 0.89 & 2.24 & 100.0 & 0.0 & 0.0 & 96.98 & 1.11 & 1.91 \\
RBF-SVM & P & 100.0 & 0.0 & 0.0 & 20.3 & 19.1 & 60.6 & 100.0 & 0.0 & 0.0 & 0.0 & 79.05 & 20.95 \\
 & M & 100.0 & 0.0 & 0.0 & 19.11 & 14.18 & 66.72 & 100.0 & 0.0 & 0.0 & 0.02 & 17.96 & 82.03 \\
\hline
 & N & 97.77 & 0.45 & 1.78 & 97.02 & 1.55 & 1.42 & 99.16 & 0.2 & 0.64 & 96.82 & 1.95 & 1.23 \\
KNN & P & 0.28 & 21.41 & 78.31 & 12.16 & 36.17 & 51.67 & 0.0 & 100.0 & 0.0 & 0.26 & 91.62 & 8.12 \\
 & M & 0.16 & 14.81 & 85.02 & 11.73 & 25.74 & 62.52 & 0.0 & 0.0 & 100.0 & 0.43 & 12.99 & 86.58 \\
\hline
 & N & 99.44 & 0.4 & 0.16 & 97.08 & 1.05 & 1.88 & 99.96 & 0.0 & 0.04 & 97.82 & 1.13 & 1.04 \\
DT & P & 3.98 & 92.33 & 3.69 & 9.31 & 68.77 & 21.92 & 0.0 & 100.0 & 0.0 & 0.57 & 94.96 & 4.47 \\
 & M & 0.66 & 1.82 & 97.51 & 11.74 & 15.77 & 72.48 & 0.0 & 0.0 & 100.0 & 0.5 & 4.05 & 95.45 \\
\hline
 & N & 99.75 & 0.17 & 0.08 & 97.01 & 1.22 & 1.77 & 99.94 & 0.02 & 0.04 & 98.07 & 0.89 & 1.05 \\
RF & P & 2.77 & 96.62 & 0.62 & 12.65 & 63.24 & 24.12 & 0.0 & 100.0 & 0.0 & 0.09 & 96.19 & 3.73 \\
 & M & 0.16 & 2.72 & 97.12 & 10.82 & 21.13 & 68.05 & 0.0 & 0.05 & 99.95 & 0.22 & 4.67 & 95.11 \\
\hline
 & N & 99.69 & 0.31 & 0.0 & 97.46 & 1.66 & 0.88 & 99.11 & 0.22 & 0.67 & 96.64 & 1.68 & 1.68 \\
GBDT & P & 1.16 & 96.51 & 2.33 & 4.0 & 72.0 & 24.0 & 0.0 & 99.96 & 0.04 & 0.1 & 88.43 & 11.47 \\
 & M & 3.7 & 1.23 & 95.06 & 3.65 & 25.55 & 70.8 & 0.0 & 0.0 & 100.0 & 0.0 & 6.77 & 93.23 \\
\hline
\end{tabular}}

\caption{Confusion matrices of the results of classification using the entire set of features as well as only mass and radius, both for undersampling and oversampling using Parzen window estimation (as described in Sections \ref{subsubsec:ml-algos}, \ref{subsubsec:undersampling} and \ref{subsubsec:parzen_windows}). The larger the values of the diagonal elements in the confusion matrix of a classifier, the better is the performance of the respective classifier.}
\label{table:big}

\end{table*}

The results are promising and indicate that ML can be used to effectively categorize new discoveries of planets. However, as these are general methods which can be applied to various datasets, there are no aspects to the above methods that are cut out specifically for datasets which broadly have the same nuances as the PHL-EC dataset. Hence, in the next section, we elaborate on methods that we developed with the PHL-EC dataset in mind.

\section{Novel Methods of Characterization and Classification} \label{sec:methods-novel}

\subsection{Synthetic Oversampling by Assuming a Distribution in the Data and SVM-KNN} \label{subsec:data_aug-svmknn}

Assuming a distribution in data is a common approach in different simulations in physics. For our experiments, we assume that the Surface Temperature follows a Poisson distribution \citep{Sale2015}. We randomly sample the remaining features as entire feature vectors from the original dataset; having estimated the value of surface temperature based on the distribution, we concatenate it with random feature vectors of the remaining parameters of samples from the same class. We use SVM-KNN \citep{Peng2013} to ensure the purity in the class-belongingness of the synthetic samples generated by altering or rectifying the class of the synthetic samples which represent a different class. The steps in the algorithm are as follows:

\newcounter{scountersvmknn}
\begin{list}{{\bf Step \arabic{scountersvmknn}}:~}{\usecounter{scountersvmknn}}
\item The best boundary between the psychroplanets and mesoplanets are found using SVM with a linear kernel.
\item The distribution of either class is estimated as a Poisson distribution:

\begin{equation}
Pr(X) = \frac{e^{-\lambda}\lambda^{x}}{x!}
\end{equation}

\item Using the boundary determined in Step 1, an artificial data point is analyzed to determine if it satisfies the boundary conditions: if a data point generated for one class falls within the boundary of the respective class, the data point is kept in it's labeled class in the artificial dataset.
\item If a data point crosses the boundary of its respective class, then a K-NN based verification is applied. If 3 out of the nearest 5 neighbors belongs to the class to which the data point is supposed to belong, then the data point is kept in the artificially augmented dataset.
\item If the conditions in Steps 3 and 4 both fail, then the respective data point's class label is changed so that it belongs to the class whose properties it corresponds to better.
\item Steps 3, 4 and 5 are repeated for all the artificial data points generated, in sequence.
\end{list}

It is important to note that SVM and KNN are used here, along with density estimation, to rectify the class-belongingness (class labels) of artificially generated random samples and not as classifiers. If an artificially generated random sample is generated such that it does not conform to the general properties of the respective class (which can be either mesoplanets or psychroplanets), the class label of the respective sample is simply changed such that it may belong to the class of habitability whose properties it exhibits better. The strength of using this as a rectification mechanism lies in the fact that artificially generated points which are near the boundary of the classes stand a chance to be rectified so that they might belong to the class they better represent. Moreover, due to the density estimation, points can be generated over an entire region of the feature space, rather than augmenting based on individual samples. This aspect of the simulation is the cornerstone of the novelty of this approach: in comparison to existing approaches as SMOTE (Synthetic Minority Oversampling Technique) \citep{Chawla:2002:SSM:1622407.1622416}, the oversampling does not depend on individual samples in the data.

\subsection{Elastic Splitting Criteria for Decision Trees} \label{subsec:egini}

An interesting observation that can be made from the data is that the non-habitable class of exoplanets has a more expansive distribution in the feature space while the classes of mesoplanets and psychroplanets occupy a small range. This bias is more difficult to handle than a class-proportion bias as a small variance in some classes could easily be confused with the more distributed class, or be less representative to a classifier. Traditionally, splitting criteria for decision trees do not account for this and provide an equitable representation to all classes. However, taking into consideration structure and spread of the data, we try to assign elastic exponential parameters to all the probabilities in the Gini impurity index to develop an asymmetric representation for a split \citep{Zighed2010}. Our intention is to adjust the bias towards the mesoplanet and psychroplanet classes so that the final results obtained are more efficacious.

\begin{figure*}
\begin{center}
\begin{subfigure}[t]{0.45\columnwidth}
	\includegraphics[width=\columnwidth]{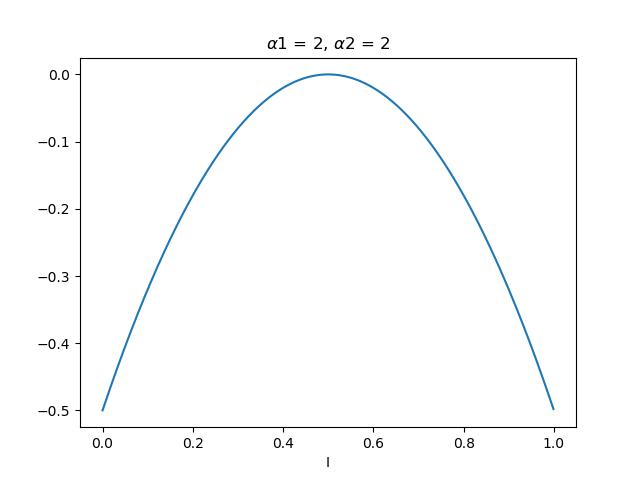}
	\caption{This is a visualization of the standard Gini impurity criteria. It is a perfectly symmetric function on two probability variables: there is no preference or asymmetry of having better classifications towards any class in the dataset. The highest impurity encountered here is when both classes have an equal number of samples in the same node, i.e., the proportion of each class is $50\%$.}
\end{subfigure}
~
\begin{subfigure}[t]{0.45\columnwidth}
	\includegraphics[width=\columnwidth]{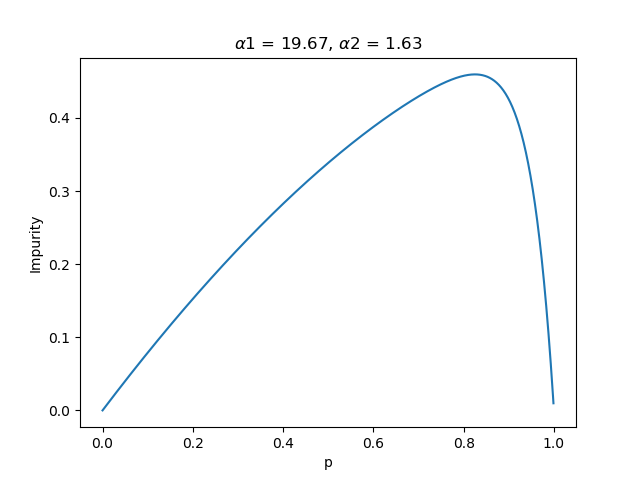}
    \caption{Here we see that the function has been skewed. The implication of this is that impurity due to the class $c_1$ with probability $p$ is less for a large range of $p$ as compared to the standard Gini impurity. As a consequence of the peak being shifted towards the right, the maximum impurity of $c1$ comes from a greater number of samples than what results in a $50\%$ proportion.}
\end{subfigure}

\begin{subfigure}[t]{0.45\columnwidth}
	\includegraphics[width=\columnwidth]{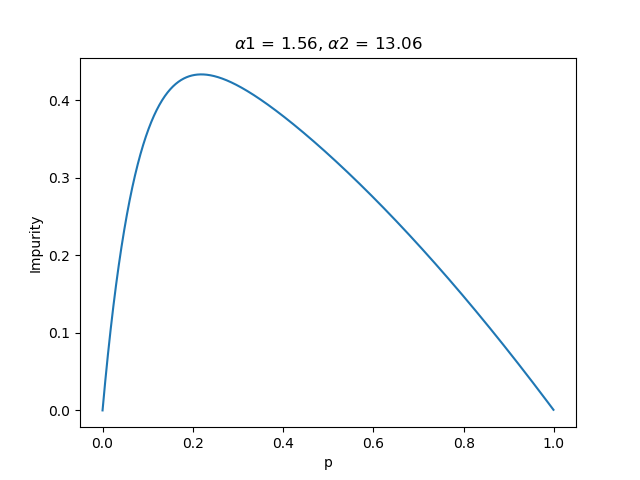}
	\caption{This is similar to the case in Figure \ref{fig:splitting-criteria-viz} (b), except that the impurity that results from the class with probability $p-1$ is lesser.}
\end{subfigure}
~
\begin{subfigure}[t]{0.45\columnwidth}
	\includegraphics[width=\columnwidth]{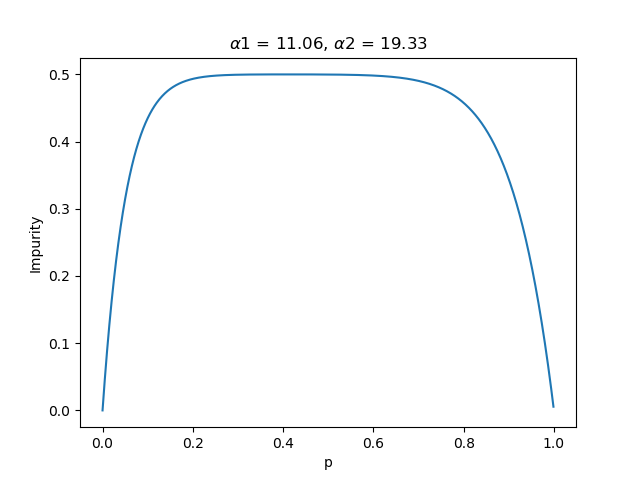}
% * <suryodaybasak@gmail.com> 2018-05-08T08:01:10.004Z:
%
% ^.
	\caption{This is another variant that can arise due to the exponents ($\alpha$s) when both classes contribute proportionately, and neither gets a clear preference: confusion may arise and such cases should be avoided with appropriate parameter tuning.}
\end{subfigure}

\caption{All the graphs plotted here are for a 2-class classification problem. If the probability of occurrence of a sample of one class in a node is $p$, then consequently, the probability of occurrence of a sample of the other class in the same node is $1-p$.}

\label{fig:splitting-criteria-viz}
\end{center}

\end{figure*}

The elastic splitting criteria is given by:

\begin{equation}
\begin{split}
I(p_1, p_2, ... , p_n) &= k \Big(1 - \sum_{c = 1}^{n}p_{i}^{\alpha_{i}}\Big)\\
G &= I_{N} - I_{L} - I_{R}
\end{split}
\end{equation}

where $I$ is the impurity of a node \citep{Quinlan1986}, $n$ is the number of classes in the dataset, $p_i$ is the probability of finding an instance of the $i^{th}$ class in a node, $\alpha_{i}$ is the elasticity associated with the $i^th$ class in the data. In order to ascertain the best split in a node in a decision tree, the gain $G$ is used and $I_N$, $I_L$ and $I_R$ represent the impurity of the parent node and potential left and right child nodes after a split, respectively. In our implementation, we have taken $k = 5$ so that this criteria is backward compatible with other existing libraries. The new addition to this splitting criteria are the elasticities $\alpha_i$ which are constants that \textit{skew} the response of the splitting criteria based on our preference and are supplied to the algorithm as parameters. The skewed response is visualized in Figure \ref{fig:splitting-criteria-viz}. What is clear is that for any $\alpha_i > 1$, the functional form of $I$ is concave, which is a condition for a function to be used as a splitting criteria \citep{Breiman1996}. The first and second partial derivatives of $I(p_1, p_2, ... , p_n)$ are:

\begin{equation}
\begin{split}
\frac{\partial I}{\partial p_i} &= -\alpha_i p_{i}^{\alpha_i - 1}\\
\frac{\partial I}{\partial p_j\partial p_i} &= \begin{cases}
-\alpha_i(\alpha_i - 1) p_{i}^{\alpha_i - 2},& \text{if } i = j\\
0,& \text{if } i \neq j\\
\end{cases}
\end{split}
\end{equation}

The Hessian matrix of $I(p_1, p_2, ... , p_n)$ consequently is:

\begin{equation}
H(I(p_1, p_2, ... , p_n)) = \begin{bmatrix}
    -\alpha_1(\alpha_1 - 1) p_{1}^{\alpha_1 - 2} & 0 & \dots & 0\\
    0 & -\alpha_2(\alpha_2 - 1) p_{2}^{\alpha_2 - 2} & \dots & 0\\
    \vdots & \vdots & \ddots & \vdots\\
    0 & 0 & \dots & -\alpha_n(\alpha_n - 1) p_{n}^{\alpha_n - 2}\\
\end{bmatrix}
\end{equation}

We briefly justify the concavity from the Hessian matrix by dealing with a 3-class scenario as this is relevant to our work. For a function to be concave, its Hessian matrix must be symmetric and negative definite. This implies that the odd numbered primary minors must be negative and the even numbered primary minors must be positive . Let $D_i$ represent the $i^{th}$ primary minor. In the Hessian matrix of the elastic Gini, we have:

\begin{equation}
\begin{split}
D_1 &= -\alpha_1(\alpha_1 - 1) p_{1}^{\alpha_1 - 2}\\
D_2 &= \alpha_1(\alpha_1 - 1) p_{1}^{\alpha_1 - 2} \times \alpha_2(\alpha_2 - 1) p_{2}^{\alpha_2 - 2}\\
D_3 &= -\alpha_1(\alpha_1 - 1) p_{1}^{\alpha_1 - 2} \times \alpha_2(\alpha_2 - 1) p_{2}^{\alpha_2 - 2} \times \alpha_n(\alpha_n - 1) p_{n}^{\alpha_n - 2}
\end{split}
\end{equation}

$D_1$ and $D_3$ can be negative and $D_2$ can be positive iff $\alpha_i > 1 \forall i$. Thus, following this condition, the function is concave, and we ensure in our computer program that we supply elasticities that are strictly greater than 1.

We implemented this method using the \textit{anytree} module in the Python programming language running on a Linux-based OS on a computer with a 2.2GHz, dual-core, Core-i3 processor.

\subsection{Understanding Surface Temperature Based Discrimination of Exoplanets Based on Machine Learning} \label{subsec:tss}

After ascertaining the effectiveness of ML algorithms for the automatic classification of exoplanets, we developed a method to quantify the potential of a planet to be habitable, based on surface temperature alone, by developing a metric which is entirely data-driven. We call this the Thermal Suitability Score (TSS) because we develop it by using the mean surface temperature of a planet (and features extracted from the surface temperature)

\subsubsection{Thermal Suitability Score}

The Thermal Suitability Score (TSS) is a score which, in addition to providing a notion of similarity to Earth in terms of surface temperature, provides a habitability classification of an exoplanet. The formulation of this method is based on SVMs. As a part of this method, two classes are used based on the optimistic sample of potentially habitable exoplanets by PHL \citep{phlref}. The two classes are those of \textit{potentially habitable} and \textit{non-habitable} exoplanets. The TSS is determined by first finding the maximum separating \textit{hyperplane} between the classes in the data, which acts as a discriminator, and using the distance from the hyperplane as the key characteristic. The metric is then developed by normalizing this distance by dividing by the distance of the Earth's feature vector from the hyperplane.

The goal of this model is a to find score that can instantly help us discriminate between potentially habitable and non-habitable planets by finding one boundary between two classes in the data. This is a hybrid approach where a model outputs a number and a sign, the number indicating similarity to earth and the sign indicating the class. In this light, Surface Temperature is one of the only features which can be used to develop the metric because S. Temp (and the related features of flux and distance from parent star) are the only features based on which the \textit{habitable} and \textit{non-habitable} samples are \textit{reasonably} linearly separable.

%The data used to develop and test this metric comprise of two classes, the first being \textit{optimistic habitable sample} as proposed by PHL \textbf{cite}, as the habitable class, and the second, being the compliment of the habitable class as the non-habitable class. There exists a small overlap between the two classes near the boundaries (\textbf{insert figure ref}).

\subsubsection{Formulation of the Optimization Problem}

The SVM quadratic optimization problem \citep{vapnik1964}, which is the basis of the TSS, is given as:

\begin{equation}
\begin{split}
\!\min_{\lambda} \hspace{0.2cm} & \frac{1}{2} \lambda^{T} (\textrm{y}\textrm{y}^{T}K)\lambda - \lambda \\
\text{subject to} \hspace{0.2cm} & -\lambda \leq 0, \\
 & \textrm{y} \cdot \lambda = 0
\end{split}
\end{equation}

where $\textrm{y}$ is the list of class labels corresponding to samples in the data, $\lambda$ is the set of Lagrange multipliers, and $K$ is the Gram matrix, which is given as:

\begin{equation}
K(\textrm{x}_1, ... , \textrm{x}_m) = \begin{bmatrix}
    \textrm{x}_{1} \cdot \textrm{x}_{1} & \textrm{x}_{1} \cdot \textrm{x}_{2} & \dots & \textrm{x}_{1} \cdot \textrm{x}_{m}\\
    \textrm{x}_{2} \cdot \textrm{x}_{1} & \textrm{x}_{2} \cdot \textrm{x}_{2} & \dots & \textrm{x}_{2} \cdot \textrm{x}_{m}\\
    \vdots & \vdots & \ddots & \vdots\\
   \textrm{x}_{3} \cdot \textrm{x}_{1} & \textrm{x}_{3} \cdot \textrm{x}_{2} & \dots & \textrm{x}_{3} \cdot \textrm{x}_{m}\\
\end{bmatrix}
\end{equation}

where $\textrm{x}_i$ represents the $i^{\textrm{th}}$ sample in the data.

After the optimization problem has been solved and the support vectors have been found, the weight and bias: the variables $\textrm{w}$ and $b$ are determined by:

\begin{equation}
\begin{split}
\textrm{w} &= \sum_{i = 1}^{m} \lambda_{i}y_{i}\textrm{x}_i \\
b &= \frac{1}{m} \sum_{i = 1}^{m} y_i - \textrm{w} \cdot \textrm{x}_i \\
\end{split}
\end{equation}
 where $m$ is the number of samples in the dataset.

The features used in this method are of the following form:

\begin{equation}
\textrm{x} = (T, |T-1|)
\end{equation}

where $|\cdot|$ represents the absolute value function and $T$ is the surface temperature in Earth units. Together, these two features give us a data representation of the surface temperature and the similarity of the surface temperature of the planet to that of the Earth's. This implies that the Earth's feature vector is $(1,0)$ and the consequence of this is that in the feature space, the distance of Earth from the maximum separating hyperplane is the maximum. In addition to this, as a consequence of the discrimination done by the hyperplane, the output of the method is positive for potentially habitable samples (and non-habitable samples whose surface temperatures are near the hyperplane) and it is negative for non-habitable samples. Let the distance of the Earth from the maximum separating hyperplane be represented as $d$. The final expression for the score is thus given as:

%In this model, we use the distance of Earth from the hyperplane as a scaling factor; we divide the distance of any planet's feature vector from the maximum separating hyperplane by the distance of Earth from the maximum separating hyperplane, represented as $d$. This number thence achieved is used as a score which indicates habitability numerically based on surface temperature of an exoplanet. The final expression for the score is thus given as:

\begin{equation}
TSS = \frac{y \cdot \big(\textrm{w} \cdot \textrm{x} + b \big)}{d}
\end{equation}

\subsubsection{Feature Extraction: What are the Model Parameters?}

We use a parameter in addition to the value of the surface temperature of the exoplanets: the absolute value of the difference between the surface temperature of the exoplanet and the surface temperature of the Earth, whose value in EU is 1. Thus, the model inputs become ordered pairs of the type $(T, abs(T - 1))$. A data implication of extracting a feature this way is that the value of $abs(T - 1)$ is $0$ for Earth, an aspect central to the scoring mechanism of the model.

\begin{figure}[htbp!]
\begin{center}
\includegraphics[width=0.7\columnwidth]{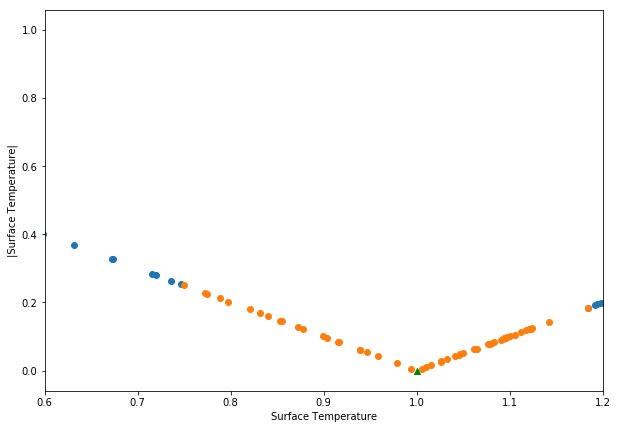}
\caption{Plot of S. Temp vs absolute value of (S. Temp - 1). In EU, 1 is the S. Temp of Earth, and in this graph, it is represented by the green triangle at $(1,0)$. The points in orange represent the optimistic sample of potentially habitable exoplanets and the points in blue represent non-habitable planets. Since the non-habitable set is more expansive, only the points in the vicinity of the habitable samples are plotted. We see that there is minor overlap near the boundaries of the classes.}
\label{fig:temp-abs}
\end{center}
\end{figure}

From a physical viewpoint, we now have a representation of a planet's surface temperature in comparison to Earth. From a computational viewpoint, we have an added dimension in the dataset which will help us effectively separate and score the potentially habitable planets from the non-habitable planets using a single hyperplane in a 2D space.

\subsubsection{Overlap Between Classes}

In the SVM formulation for a linearly inseparable dataset \citep{vapnik1964}, there is a minimization of a classification error which provides the \textit{best} boundary between the classes in the data. However, in this method, we do not want to find a best-case boundary of separation, but would like to be inclusive of the habitable samples which are manually labeled by the PHL-EC as we consider them to be reliable points of judgment of habitability. Hence, we find the convex hull of the habitable samples and exclude the non-habitable samples within this convex hull prior to finding a separating hyperplane. By doing this, we get a perfect separation between the two classes, and we proceed to find the optimal separating hyperplane. %Should there be any misclassifications, it will be of a confusion of the form where we know that the error of misclassification of habitable planets is zero, but 

\begin{figure}[htbp!]
\begin{center}
\includegraphics[width=0.7\columnwidth]{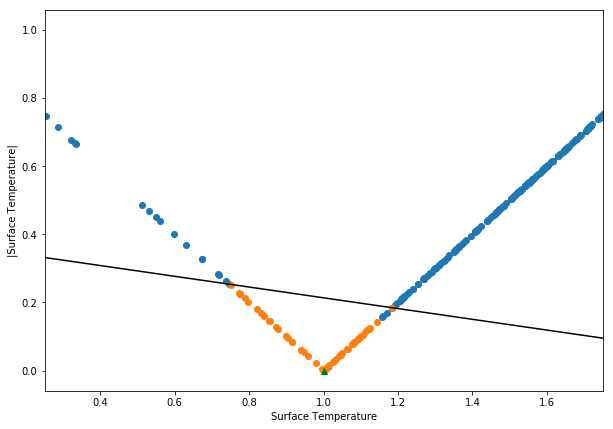}
\caption{A graphical depiction of the optimal separating hyperplane (black line) determined after disregarding the non-habitable points within the convex hull of samples of the habitable class. The consequence of this is noticed near the class boundaries, where a few non-habitable samples fall on the habitable-side of the hyperplane. }
\label{fig:temp-abs-hyp}
\end{center}
\end{figure}

\subsubsection{A Geometric Interpretation of the Method and Proof of Maximum Value}

Let the feature vectors be denoted as $(T, g(T))$, where

\begin{equation}
  g(x)=\begin{cases}
    L_1 = -T + 1, & \forall T < 1\\
    L_2 = T - 1, & \forall T > 1
  \end{cases}
\end{equation}

g(T) is nothing but an expansion of the absolute value function. $L_1$ and $L_2$ may be considered as two lines which intersect at $(0,1)$. Let the separating hyperplane be denoted by $H$. As we know that a separating hyperplane in a 2D space is a line, $L_1$, $L_2$ and $H$ may be considered to form a triangular region if the angle made b $H$ with respect to $L_1$ and with respect to $L_2$ is zero. This is proven below.

Let the angle between $H$ and $L_1$ and $L_2$ be $\theta_1$ and $\theta_2$ respectively. If $\textrm{sin}\theta_1 > 0$ and $\textrm{sin}\theta_2 > 0$, then we can say that $H$ is not parallel or collinear with respect to $L_1$ and $L_2$ respectively. If $\textrm{sin}\theta_1$ and $\textrm{sin}\theta_2$ are both greater than $0$ then $H$ intersects with both $L_1$ and $L_2$.

The slope of $L_1$ is $-1$ and the slope of $L_2$ is $1$. Let the angle between $L_1$ and $L_2$ be given by $\theta_3$. Then,

\begin{equation}
\begin{split}
\theta_3 &= \textrm{arctan}\frac{-1 -(1)}{1 + (-1)(1)}\\
&= \textrm{arctan} \frac{-2}{0}\\
&= \frac{\pi}{2}
\end{split}
\label{eq:rt_angle}
\end{equation}

Thus, considering the points of intersection of $H$ with $L_1$ and $L_2$ being $A = (x_1, y_1)$ and $B = (x_2, y_2)$, and considering $O = (1,0)$, we can assert that a triangular region is formed by $OAB$. Also, from Equation \ref{eq:rt_angle}, we know that $\bigtriangleup OAB$ is a right angled triangle.

\begin{figure}[htbp!]
\begin{center}
\includegraphics[width=0.5\columnwidth]{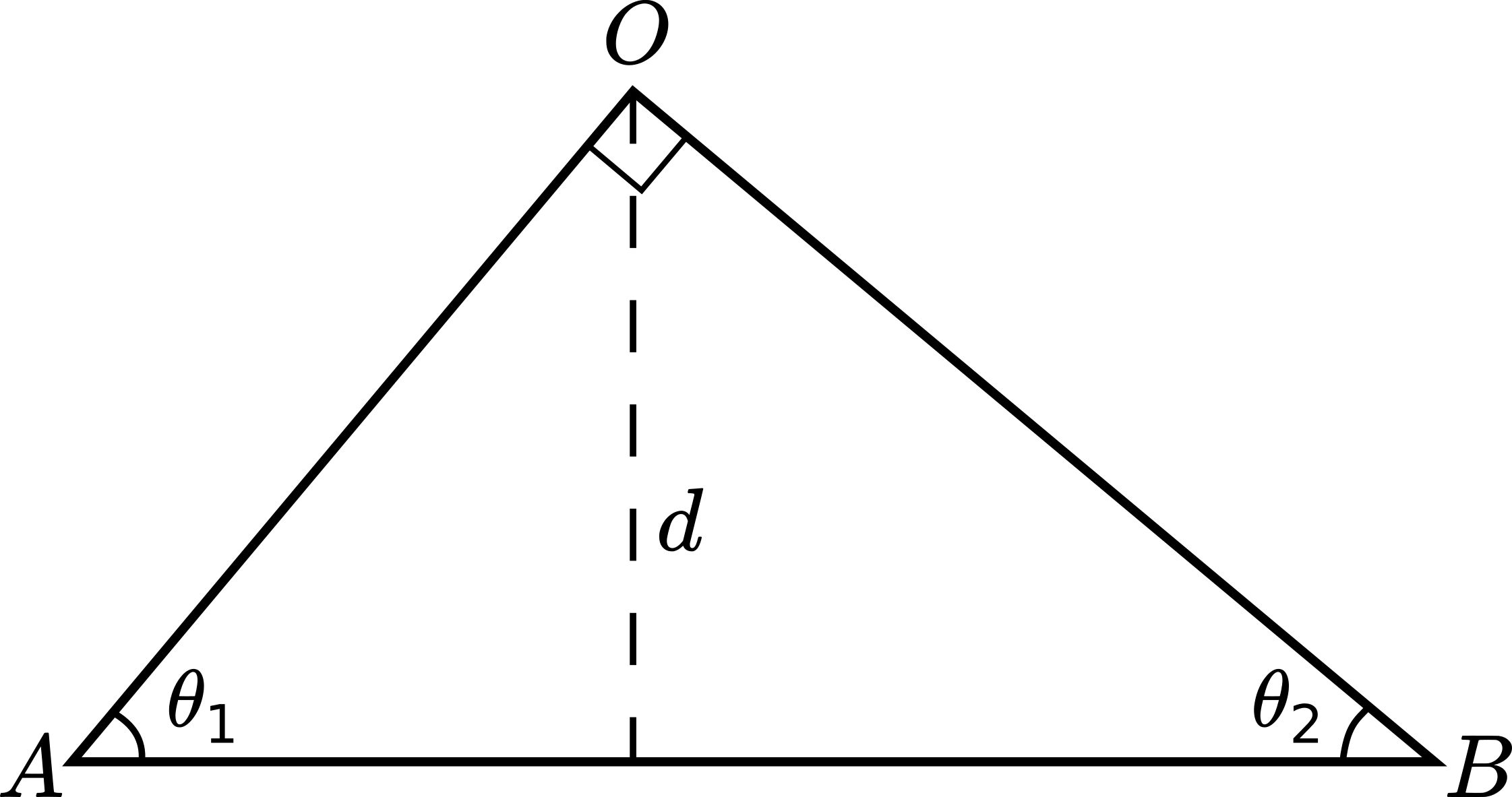}
\caption{Geometric representation of the problem in a 2D space. $AB$ is a segment from the optimal separating hyperplane, $A = (x_1, y_1)$, $B = (x_2, y_2)$ and $O = (1,0)$. $d$ is the distance of $(1,0)$ (which is Earth's feature vector) from the separating hyperplane.}
\label{fig:rt_angle}
\end{center}
\end{figure}

Let us consider the side $OA$. Here, $O$ and $A$ are the end points. On this line segment, we know that point $O$ is the greatest distance away from point $A$. This is proved by contradiction.

Let us assume that on this line segment, $O$ is not at the greatest distance away from A. Let there be a point $O'$ on $OA$ such that it lies between $O$ and $A$ and is at a greater distance away from $A$ than $O$. We know that for any point $K$ between $O$ and $A$,

\begin{equation}
|OA| = |AK| + |KO|
\end{equation}

Hence, for the point $O'$,

\begin{equation}
\begin{split}
|AO| & = |AO'| + |O'O| \\
\Rightarrow |AO'| &= |AO| - |O'O| \\
\Rightarrow |AO'| &< |AO|
\end{split}
\end{equation}

However, this contradicts the premise that there can be a point $O'$ on the line segment $OA$ for which $|O'A| > |OA|$. This implies that on line segment $OA$ the greatest distance between any two points on the line is the distance between the endpoints $O$ and $A$ and this is the length of the line. This further implies that there can be no point on $OA$ apart from $O$ for which $|A'O|\textrm{sin}\theta_1 > |AO|\textrm{sin}\theta_1$.

Keeping in mind the geometric representation as shown in Figure \ref{fig:rt_angle}, we see that the distance between $O$ and $AB$, which actually represents the segment of the hyperplane included in the triangular region $\bigtriangleup OAB$, is given by $|AO|\textrm{sin}\theta_1$. The same can be proven by taking into consideration side $OB$ instead of $OA$.

Thus, in the context of the problem, in the feature space, Earth is the furthest away from the maximum separating hyperplane and the distance of any planet with feature vector not equal to $(0,1)$ from $H$ will be less than that of Earth.

%In this model, we use the distance of Earth from $H$ as a scaling factor; we divide the distance of any planet's feature vector from the maximum separating hyperplane by the distance of Earth from the maximum separating hyperplane, given by $d$. This number thence achieved is used as a score which indicates habitability numerically based on surface temperature of an exoplanet. The final expression for the score is thus given as:

%\begin{equation}
%TSI = y\Bigg(\frac{\textrm{w}}{\|\textrm{w}\|}\cdot \textrm{x} %+ \frac{b}{\|\textrm{w}\|} \Bigg)
%\end{equation}

%\begin{equation}
%TSI = \frac{y \cdot \big(\textrm{w} \cdot \textrm{x} + b \big)}{d}
%\end{equation}

Thus, finally, the conditions that arise which allow this model to be used as a metric is that the separating hyperplane should not be collinear or parallel to any of the sides of the triangular region formed by $\bigtriangleup OAB$. While solving the problem, we find that this condition is satisfied by the data. We find $\textrm{w} = [-392.011, -2487.989]$, $b = 923.011$ and $d = 531.0002$. The solution of the problem was programmed in Python3.6 with the library CVXOPT, which is a library for convex optimization.

%\FloatBarrier  

\subsection{Results of New Methods of Classification} \label{subsec:results-classif-novel}

These results are presented in the same way as in Table \ref{table:big}, with confusion matrices for classification methods. Additionally, we present the results of TSS by presenting a representative sample of the scores.

\begin{enumerate}

\item \textbf{Synthetic Oversampling Using Distribution Assumption and SVM-KNN}: Similar to the method of oversampling using KDE, 1000 samples each for the classes of mesoplanets and psychroplanets were artificially generated. The classifiers described in Section \ref{subsec:methods-existing} are used after the data generation.

\item \textbf{Decision Trees and Random Forests with the Elastic Gini Splitting Criteria}: We incorporate the elastic Gini criteria into decision trees and random forests and present the results in the Table \ref{tab:egini-results}.

\item \textbf{Values of Thermal Suitability Scores}: We present the TSS of a sample of potentially habitable and non-habitable exoplanets in order to compare how the metric behaves for different exoplanets, and to understand the relevance of the scores. The samples of TSS is presented in Table \ref{tab:tss}.

\end{enumerate}

\begin{table*}[!htbp]
\centering
\begin{tabular}{ | c | c | c c c | c c c | }

\hline

%& & \multicolumn{3}{c}{All Features} & \multicolumn{3}{c}{Mass and Radius only}\\
\multirow{2}{*}{Algorithm} & \multirow{2}{*}{\backslashbox{True}{Pred}} & \multicolumn{3}{c|}{All Features} & \multicolumn{3}{c |}{Mass and Radius Only}\\ \cline{3-8}

%\hline
 &  & N & P & M & N & P & M\\
\hline

 & N & 99.72 & 0.12 & 0.16 & 96.79 & 0.82 & 2.39 \\
Gaussian Naive Bayes & P & 0.0 & 100.0 & 0.0 & 0.0 & 15.44 & 84.56  \\
 & M & 0.0 & 0.0 & 100.0 & 0.0 & 0.51 & 99.49  \\

\hline
 & N & 99.12 & 0.13 & 0.75 & 94.29 & 2.11 & 3.6 \\
Linear Discriminant Analysis & P & 0.0 & 93.01 & 6.99 & 0.0 & 14.41 & 85.59 \\
 & M & 0.0 & 0.0 & 100.0 & 0.0 & 6.7 & 93.3 \\

\hline
 & N & 99.61 & 0.16 & 0.23 & 96.29 & 1.8 & 1.91  \\
Support Vector Machine & P & 0.0 & 100.0 & 0.0 & 0.0 & 73.24 & 26.76 \\
 & M & 0.0 & 0.0 & 100.0 & 0.0 & 56.5 & 43.5 \\

\hline
 & N & 100.0 & 0.0 & 0.0 & 97.17 & 1.14 & 1.69 \\
Radial Basis SVM & P & 0.0 & 100.0 & 0.0 & 0.0 & 92.54 & 7.46 \\
 & M & 0.0 & 0.0 & 100.0 & 0.0 & 18.11 & 81.89 \\

\hline
 & N & 99.14 & 0.21 & 0.65 & 97.09 & 0.96 & 1.94 \\
K Nearest Neighbors & P & 0.0 & 100.0 & 0.0 & 0.0 & 98.91 & 1.09  \\
 & M & 0.0 & 0.0 & 100.0 & 0.0 & 4.84 & 95.16  \\

\hline
 & N & 99.97 & 0.0 & 0.03 & 98.32 & 0.74 & 0.93  \\
Decision Trees & P & 0.0 & 100.0 & 0.0 & 0.0 & 99.77 & 0.23 \\
 & M & 0.0 & 0.0 & 100.0 & 0.0 & 5.41 & 94.59 \\

\hline
 & N & 99.95 & 0.02 & 0.03 & 98.36 & 0.63 & 1.01 \\
Random Forests & P & 0.0 & 100.0 & 0.0 & 0.0 & 99.7 & 0.3 \\
 & M & 0.0 & 0.0 & 100.0 & 0.0 & 5.45 & 94.55 \\
\hline

 & N & 99.16 & 0.0 & 0.84 & 96.57 & 1.22 & 2.21 \\
GBDT & P & 0.0 & 100.0 & 0.0 & 0.0 & 100.0 & 0.0  \\
 & M & 0.0 & 0.0 & 100.0 & 0.0 & 5.68 & 94.32 \\
\hline
\end{tabular}

\caption{Confusion matrices of the results of classification using the SVM-KNN oversampling technique described in Section \ref{subsec:data_aug-svmknn} based on the entire feature set and only mass and radius as features.}
\label{tab:smol}
\end{table*}

\begin{table*}[!htbp]
\centering
\begin{tabular}{| c | c | c c c | c c c |}

\hline

%& & \multicolumn{3}{c}{All Features} & \multicolumn{3}{c}{Mass and Radius only}\\
\multirow{2}{*}{Algorithm} & \multirow{2}{*}{\backslashbox{True}{Pred}} & \multicolumn{3}{c|}{All Features} & \multicolumn{3}{c|}{Mass and Radius Only}\\ \cline{3-8}

%\hline
 &  & N & P & M & N & P & M\\
\hline

 & N & 97.99 & 0.0 & 2.01 & 98.98 & 0.0 & 1.02 \\
Decision Trees with Balancing & P & 0.0 & 100.0 & 0.0 & 3.22 & 87.10 & 9.68  \\
 & M & 0.0 & 0.0 & 100.0 & 0.0 & 1.82 & 98.18  \\

\hline
 & N & 94.18 & 4.76 & 1.05 & 94.27 & 0.29 & 5.44 \\
Decision Trees Without Balancing & P & 0.0 & 66.67 & 33.33 & 0.0 & 50.0 & 50.0 \\
 & M & 0.0 & 0.0 & 100.0 & 0.0 & 0.0 & 100.0 \\
\hline
\end{tabular}

\caption{Confusion matrices of the results of classification of decision trees with the elastic Gini criteria (elaborated in \ref{subsec:egini}). Compared to the results shown in Tables \ref{table:big} and \ref{tab:smol}, these results are better, affirming the idea of handling asymmetry in the dataset explicitly.}
\label{tab:egini-results}
\end{table*}

\subsection{Results of Thermal Suitability Score Function} \label{subsec:results-tss}

For the sake of clarity, we have included two sets of TSS values (shown in Table \ref{tab:tss}), one for potentially habitable exoplanets, and the other for non-habitable exoplanets as we have proven that the values of the scores can be positive and negative facilitating perfect discrimination between classes.

\begin{table}
\centering
\begin{tabular}{| c c c | c c c |}
\hline
\multicolumn{3}{|c|}{Potentially Habitable Exoplanets} & \multicolumn{3}{c|}{Non-Habitable Exoplanets}\\
\hline
P. Name & S. Temp & TSS & P. Name & S. Temp & TSS\\
\hline
TRAPPIST-1 d & 1.01527 & 0.91713 & TRAPPIST-1 b & 1.37674 & -1.04331\\
TRAPPIST-1 e & 0.90416 & 0.62172 & TRAPPIST-1 c & 1.20799 & -0.12806\\
TRAPPIST-1 f & 0.79757 & 0.20096 & TRAPPIST-1 h & 0.63125 & -0.45554\\
TRAPPIST-1 g & 0.75035 & 0.01456 & Kepler-519b & 1.97257 & -4.27495\\
ProximaCenb & 0.91632 & 0.66969 & MOA-2010-BLG-328Lb & 0.33403 & -1.62874\\
Kepler-186f & 0.77430 & 0.10913 & OGLE-2005-390Lb & 0.32187 & -1.67671\\
Kepler-705 b & 1.00555 & 0.96987 & Wolf1061d & 0.56042 & -0.73513\\
K2-72e & 1.10555 & 0.42749 & YZCetb & 1.64861 & -2.51789\\
Ross128b & 1.09410 & 0.48964 & GJ649c & 1.98403 & -4.33710\\
K2-3d & 1.14271 & 0.22599 & EPIC211822797b & 1.44305 & -1.40301\\
\hline
\end{tabular}
\caption{The TSS (\ref{subsec:tss}) of various samples are presented. The potentially habitable sample in this table mostly consists of exoplanets that have been gaining a lot of popularity as potentially habitable worlds. The non-habitable samples are mostly chosen at random, except the TRAPPIST-1 planets, which we included for the sake of completeness with respect to the TRAPPIST-1 potentially habitable samples. Consider the planets TRAPPIST-1 d and b. The differences in sign indicate clear demarcation between the two different classes of habitability. In stark contrast, S.Temp based classification shall place TRAPPIST-1 d and b in the same class because of the proximity of the decision boundary (both  TRAPPIST-1 d and b having the same sign as well as close in magnitude). The TSS, unlike S.Temp can thus bolster the discrimination capability (change in sign generates a non-ambiguous separating hyperplane) of a habitability classifier. The variation of the scores from TRAPPIST-1 b to h are reflective of the knowledge gained from ongoing research on the TRAPPIST-1 planets \citep{Barr2017,deWit2018}}.
\label{tab:tss}
\end{table}

\section{Discussion} \label{sec:discussion}

\subsection{The need for Training Classifiers on Balanced Datasets} \label{subsec:about-balancing}

Predominantly in the case of metric classifiers, an imbalanced training set can lead to misclassification. The classes which are underrepresented in the training set might not be classified as well as the dominating class. In the PHL-EC dataset, the non-habitable category is over 1000 times as large in terms of the number of samples compared to the mesoplanet and psychroplanet classes. Upon inspection of the confusion matrices of the classification done using all the features and the datasets without balancing, we can see that the results are biased and almost every sample will be classified as a non-habitable sample.

\subsection{Order of Importance of Features} \label{subsec:feature-imp}

In any large dataset, it is natural for certain features to contribute more towards defining the characteristics of the entities in that set. In other words, certain features contribute more towards class belongingness than certain others. As a part of the experiments, the we wanted to observe which features are more important. The ranks of features and the percentage importance for random forests and for GBDTs (using XGBoost) are presented in Table \ref{tab:feat-imp}. Every classifier uses the features in a dataset in different ways. That is why the ranks and percentage importances observed using random forests and GBDTs are different. The feature importances were determined using artificially balanced datasets.

\begin{table}
\centering
\begin{tabular}{| c c | c c |}
\hline
\multicolumn{2}{|c|}{Random Forests} & \multicolumn{2}{c|}{GBDT/XGBoost}\\
\hline
Feature & Importance \% & Feature & Importance \% \\
P. Habitable & 14.7 & P. Ts Min (K) & 31.43 \\
P. Ts Mean (K) & 12.47 & P. Habitable & 28.57 \\
P. Ts Min (K) & 10.54 & P. Zone Class & 14.29 \\
P. ESI & 8.72 & P. Min Mass (EU) & 5.71 \\
P. Ts Max (K) & 6.91 & P. Density (EU) & 5.71 \\
P. HZI & 6.57 & P. HZD & 5.71 \\
P. Inclination (deg) & 5.33 & P. Omega (deg) & 5.71 \\
P. Min Mass (EU) & 3.43 & P. HZC & 2.86 \\
P. SFlux Mean (EU) & 3.22 & -- & -- \\
P. Teq Mean (K) & 3.22 & -- & -- \\
\hline
\end{tabular}
\caption{The feature importances with respect to random forests and GBDTs are presented. We have presented the top 10 features for random forests; GBDTs selected 8 features and the ranking is provided. There are important similarities between the most important features as per each method, disregarding the order. The feature ranks demonstrate that the ML algorithms are in agreement with our knowledge of physics.}
\label{tab:feat-imp}
\end{table}

\subsection{Results Using Different Feature Sets} \label{subsec:diff-feat-sets}

In Tables \ref{table:big}, \ref{tab:smol} and \ref{tab:egini-results} where we have presented the results of all the classification runs, we see that the accuracies of classifiers on the restricted feature set of mass and radius only are generally lower compared to when the entire feature set is used. As the different classes of habitability are based on thermal properties, naturally, surface temperature emerges as the most important feature. Thus, most of the classifiers work well when all the features are used. However, when only the fundamental features of mass and radius are used to train the classifiers, the accuracies fall. Considerably robust are the tree-based classifiers of decision trees, random forests and gradient boosted decision trees, whose performance is good even when the restricted feature set is used.

Another interesting observation is that the classifiers are better at classifying the non-habitable samples. This is more pronounced when only mass and radius are used as features. Geometrically, mesoplanets and psychroplanets occupy a narrow band of values along every feature, especially surface temperature. The features of non-habitable samples comprise of a larger range of values. This is why a probabilistic classifier such as Gaussian Na\"{i}ve Bayes', or hard-boundary classifiers SVM and LDA, and instance-based classifier KNN make a lot of wrong predictions when the restricted feature set is used. The non-habitable class is more distributed in the feature space, and an appropriate proportion of samples is required to ascertain correct classifications.

\subsection{Results After Artificially Augmenting Data Samples} \label{subsec:after-data-aug}

The artificial data samples provide a more representative distribution of samples of each class. Naturally, the performance of all the classifiers are better.

In particular, the greatest increase in the accuracies is that of the KNN classifier. As KNN entirely depends on the geometric closeness of the test samples to training samples, with larger and less sparse training sets, the performance is improved. The performance of the tree-based classifiers is also improved and brought close to perfect.

Using a combination of undersampling of the non-habitable class and oversampling of the mesoplanet and psychroplanet classes, and the restrictive feature set of only mass and radius, we see that random forests provide the best results.

\subsection{Reason for Impressive Performance of Tree-Based Classifiers and Varying Performance of SVM} \label{subsec:tree-vs-svm}

\begin{figure}[htbp!]
\begin{center}
\includegraphics[width=0.8\columnwidth]{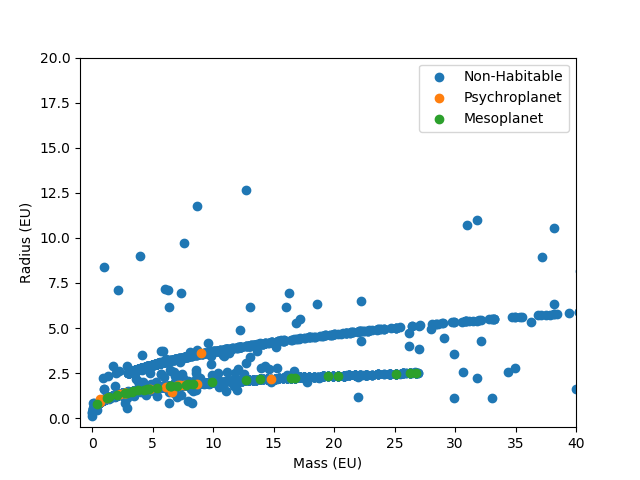}
\caption{The plot of mass vs radius shows that finding discernible trends or ranges only based on mass and radius is difficult to find. The data is cluttered and is bound to create problems for any classifier. Clearly, no simple set of hyperplanes can separate the classes and hence, SVM fails when only these two features are used. However, the tree-based classifiers perform substantially better because they partition the feature space multiple times to find different ranges where the samples geometrically fall into.}
\label{fig:m-vs-r}
\end{center}
\end{figure}

In the dataset, the classes are defined based on surface temperature. Upon training an SVM with all the features, the best boundaries between classes based on surface temperature are discovered. Upon inspection of the results, we see that SVM performs well when surface temperature is also included as a feature in the dataset. However, when SVM with a linear kernel is trained using only mass and radius, the accuracy suffers, whereas SVM with a radial basis kernel performs very poorly when the entire feature set is used, and fairly well when it is trained using mass and radius along an artificially augmented data samples. When only mass and radius are used as features, the different classes are not separated by strong boundaries and are cluttered as shown in \ref{fig:m-vs-r}. For such type of data, an SVM may not be an ideal classifier.

However, the classification of samples in the overlapping regions of the different classes can be addressed using tree-based classifiers as they employ \textit{recursive space partitioning} \citep{Quinlan1986}. This paradigm finds subspaces which correspond to different classes in the data; based on the class of the subspace a test sample geometrically falls into, the class is appropriately assigned to it. Decision trees work entirely based on this principle and random forests are an improvement over decision trees \citep{Khaidem2016} wherein subsets of all the data points and features are sampled to build multiple tree classifiers whose predictions are aggregated (this principle is known as bootstrap aggregation or \textit{bagging}). Gradient boosted decision trees (GBDT) employ further improvements over random forests where the data is modeled using a regressor function in every node of a tree learner.

It is observed that in general, under every type of experiment, the tree-based classifiers perform considerably better than the other classifiers. 

\subsection{Performance Improvement Using Elastic Gini Splitting Criteria in Decision Trees} \label{subsec:egini-awesome}

Naturally, the inquisition arises to further build up on decision trees to bolster the performance. Asymmetricity in ML approaches treats different classes with different priorities. This is often a useful thing as various critical systems may be able to afford a false positive, but not a false negative. For a scientist, stumbling upon a planet which is predicted as habitable but turns out to be non-habitable might be something that can be easily looked over, but if a planet is found to be non-habitable by a machine but in reality it is habitable, it could cost science a lot of time until the right kind of discovery is finally made.

The right paradigm of informatics sciences is to introduce to standard methods as much domain-knowledge as possible. The splitting criteria in a decision tree is at the heart of its functioning. Hence, if the criteria could address the bias directly, it would be beneficial to finding the appropriate results. By developing the eleastic Gini criteria, we could introduce to the system a mechanism to counter the sample bias in the data. The elasticities are effectively \textit{priorities} which are assigned to the system on a class-wise basis -- the lower an elasicity for a class (while keeping to the permissible range), the higher the classification priority assigned to it.

Often, accuracies are not enough to ascertain the performance of classifiers -- the numerical values of the overall accuracy using any of the methods on this dataset turned out to be extremely high. The high accuracies were a result of the sample bias -- upon an examination of the confusion matrices, it became evident that the spatial distribution and geometry played an important role in determining the right method for the task.

\subsection{Inference from the Thermal Suitability Score (TSS)} \label{subsec:inference-tss}

%One of the cornerstones of SVM as a classifier is that it is a \textit{hard-boundary} classifier, i.e. it finds a margin, or a \textit{hyperplane}, to separate classes in a dataset. Samples close to the hyperplane may be ambiguous or erroneous; the hyperplane itself may not perfectly divide the dataset into perfect class-wise partitions, but may provide a best-case discriminator.

This is a metric which is developed using ML and appropriate feature extraction. The method takes our current knowledge and uses it to discriminate and gauge the potential of incoming samples. Although optimization-based approaches have been proposed by \citep{CDHPF2016} and \citep{Proxb}, an optimization of an error function in a habitability metric has not been explored before. As it is inherently based on ML, we can increase the number of parameters as long as the notion of linear separability is maintained.

The value of this metric can only be less than 1 for all planets whose surface temperature are not equal to 1 (in EU). The consequence of this is that the value of TSS for only the Earth is equal to 1, and at this point in time, every other planet (which is a part of the PHL-EC) has a TSS of less than one. In addition to that, the \textit{hard-boundary} aspect of SVMs is used to provide results which are negative for non-habitable planets. Conclusively, the negative sign is an out-of-the-box indicator that a planet may not be thermally suitable for habitability. Samples close to the hyperplane may be ambiguous or erroneous; in this model, the hyperplane itself does not perfectly divide the dataset into perfect class-wise partitions, but provides a best-case discriminator. Some of the salient features of TSS are:

\begin{enumerate}

\item \textbf{Unidirectional Similarity Values}: The value of this metric can only be less than 1 for all planets whose S. Temp values are different from Earth. It doesn't matter if the value is greater or lesser: if it is different, then the value is below that of Earth.

\item \textbf{Positive and Negative Values}: Notice that in Table \ref{tab:tss}, in most places, the sign of the TSS has matched the CDHS class of the corresponding planet (except for one case) \citep{Proxb}, where class 5 of the CDHS \citep{} has a negative score and class 6 has a positive score. Negative represents non-habitable, and from what we know of the planets in the TRAPPIST-1 system with negative values of TSS, they're not potentially habitable.

\item \textbf{Learning from Example}: This is a metric which is developed \textit{using} ML and appropriate feature extraction. The method takes our current knowledge and uses it to discriminate and gauge the potential of incoming samples.

\item \textbf{Tackling Skewness}: From Figures \ref{fig:temp-abs} and \ref{fig:temp-abs-hyp}, we see that the distribution of the habitable samples in the feature space is not symmetric, but there exists a skewness. As a consequence of this, the separating hyperplane is not parallel to the $x$-axis. However, by thus using the separating hyperplane as a reference boundary, we can equitably judge the samples notwithstanding their respective surface temperatures being lesser than or greater than that of Earth.

\item \textbf{Scalable}: As it is inherently based on ML, we can increase the number of parameters as long as the notion of linear separability is maintained.

\item \textbf{No Planet can have an Ambiguous Score}: Technically, points on the hyperplane will have a TSS value of zero. As there are no planets which themselves lie on the maximum margin hyperplane, no planet may have a zero value.

\end{enumerate}

\section{Conclusion and Future Work} \label{sec:conclusion}

This paper presents the effectiveness of using machine learning to explore the problem of habitability classification of exoplanets. The novelty of the work lies in the appropriate exploration of the PHL-EC dataset and usage of automated classification methods that was hitherto not investigated in the existing literature. The work is a detailed investigation on exploratory data analysis involving algorithmic improvisations and machine learning methods applied to the PHL-EC dataset, bolstered by a comprehensive understanding of these methods (as documented in the appendices). The inferences drawn fortify the usage of these methods.

The dataset is unique and is a combination of both derived and fundamental planetary parameters. The derived features provide a rich representation of the exoplanets' characteristics. The accuracy of various machine learning algorithms used on the PHL-EC dataset has been computed and tabulated. Random forest, decision trees, and XGBoost perform the best with the highest class-wise accuracies closely followed by SVM; the new elastic Gini splitting criteria allows us to introduce nuances in the dataset for analysis and helps us to further improve decision trees for the task. Despite the sample bias due to the non-habitable class, we were able to achieve remarkable accuracies with classification algorithms by performing undersampling and synthetic data augmentation on the data. This goes to show that a careful study of the nature and trends of the data is a must, and simple solutions may often suffice. In addition to the implementation of classification, we have incorporated the default ideas of separability of classes in the dataset to develop the TSS which can indicate the potential habitability of an exoplanet based on the similarity of the of the exoplanet's surface temperature to the surface temperature of Earth. The outcomes of the best classifiers coupled with the feature ranking and habitability metrics (TSS as well as other metrics like CDHS) allows us to take a data-centric view of the characterization of exoplanets which can match the inferences built based on the physical study. The methods all have proofs of convergence and scalability and hence, in the future, they may be extended to incorporate newer and more relevant discoveries. As the data acquisition technology goes on improving, we will be able to incorporate new and more reliable parameters into our models and this will facilitate the efficacy of our approaches.

Exoplanets are frequently discovered and categorizing them manually is an arduous task. As data are collected, gradually adding to the volume of existing data, automatic annotation methods, along with viable strategies for discovery \citep{schulze-makuch2018time}, would eventually provide an advantage in terms of the required processing time and effort. Thus, in the future, a continuation of the present work would be directed towards achieving a sustainable and automated discrimination system for efficient and accurate analysis of different exoplanet databases. From a methodological point of view, the elastic Gini criteria can be used in decision trees, the TSS can be expanded by incorporating other features, and neural networks with activation functions inspired by the functional forms of the current methods and current data challenges can be developed for extensive analysis. We have described a novel neural network activation function in \ref{app:sbaf} inspired by our understanding of TSS. The current work, in addition to the methodological exposition, also builds on a "best-practices pipeline" for data-driven tasks.

We end this section with a note on the correct usage of ML and artificial intelligence in habitability classification of exoplanets. We do not intend to inspire any notions of automata taking over subtle and sensitive tasks such as discovering habitable worlds in the universe. Rather, we emphasize that the current work should be used to facilitate and reinforce the process of discovery and the inferences drawn be used to bolster our knowledge in the areas of exoplanet discovery.

\section{Acknowledgements} \label{sec:acknowledgements}

We would like to thank the Department of Science and Technology of India for supporting our research by providing us with resources to conduct our experiments. The project reference number is: EMR/2016/005687.

\appendix

\section{Effects of Imbalance in the Dataset} \label{app:imbalance}

In Table \ref{tab:no-balance}, we have presented the results of classification done without undersampling of the non-habitable class or synthetic augmentation of the mesoplanet and psychroplanet classes. In most of the results, there's a preference towards classification of a larger number of samples as non-habitable (after the classifier has been trained).

Thus, the methods of balancing that can be effectively used for ensuring an equitable representation of all the classes in the dataset, consequently resulting in better classification accuracies.

\begin{table*}[!htbp]
\centering
\begin{tabular}{| c | c c c c |}
\hline
Algorithm &  & Non Habitable & Psychroplanets & Mesoplanets\\
\hline

 & Non Habitable & 96.45 & 1.04 & 2.51 \\
Gaussian Naive Bayes & Psychroplanets & 6.59 & 6.59 & 86.83 \\
 & Mesoplanets & 0.0 & 6.27 & 93.73 \\
\hline

 & Non Habitable & 99.58 & 0.02 & 0.4 \\
Linear Discriminant Analysis & Psychroplanets & 100.0 & 0.0 & 0.0 \\
 & Mesoplanets & 100.0 & 0.0 & 0.0 \\
\hline

 & Non Habitable & 100.0 & 0.0 & 0.0 \\
Support Vector Machine & Psychroplanets & 100.0 & 0.0 & 0.0 \\
 & Mesoplanets & 100.0 & 0.0 & 0.0 \\
\hline

 & Non Habitable & 100.0 & 0.0 & 0.0 \\
Radial Basis SVM & Psychroplanets & 63.33 & 6.67 & 30.0 \\
 & Mesoplanets & 58.93 & 8.93 & 32.14 \\
\hline

 & Non Habitable & 100.0 & 0.0 & 0.0 \\
K Nearest Neighbors & Psychroplanets & 64.22 & 17.13 & 18.65 \\
 & Mesoplanets & 48.12 & 13.95 & 37.93 \\
\hline

 & Non Habitable & 99.48 & 0.29 & 0.23 \\
Decision Trees & Psychroplanets & 42.99 & 48.91 & 8.1 \\
 & Mesoplanets & 32.61 & 1.5 & 65.89 \\
\hline

 & Non Habitable & 99.75 & 0.12 & 0.13 \\
Random Forests & Psychroplanets & 56.86 & 33.43 & 9.71 \\
 & Mesoplanets & 32.76 & 1.72 & 65.52 \\
\hline

 & Non-Habitable & 99.95 & 0.0 & 0.05 \\
GBDT & Psychroplanets & 1.3 & 87.01 & 11.69 \\
 & Mesoplanets & 2.11 & 2.11 & 95.77 \\
\hline

\end{tabular}

\caption{Confusion matrices of the results of classification without undersampling of the non-habitable class.}
\label{tab:no-balance}
\end{table*}

\FloatBarrier

\section{Details of Data Augmentation Methods} \label{app:data_aug}

\subsection{Generating Data by Assuming a Distribution}

In order to generate 1000 samples for the classes with less number of samples (mesoplanets and psychroplanets), the hybrid SVM-KNN algorithm as described in Section \ref{subsec:data_aug-svmknn} is used to rectify the class-belongingness of any non-conforming random samples. For this, the temperature-related features of P. Ts Mean, P. Ts Min, P. Teq Min, and P. Teq Max are considered in this rectification mechanism. The artificially generated dataset is iteratively split into training and testing sets (in a ratio of 70:30). If any artificially generated sample in any iteration fails to be accepted (or be correctly classified) by the SVM-KNN algorithm, its class-belongingness in the dataset is changed. As this simulation is done only on two classes in the data, non-conformance to one class could only indicate the belongingness to the other class.

\begin{figure*}
\begin{center}
\begin{subfigure}[t]{0.45\columnwidth}
	\includegraphics[width=\columnwidth]{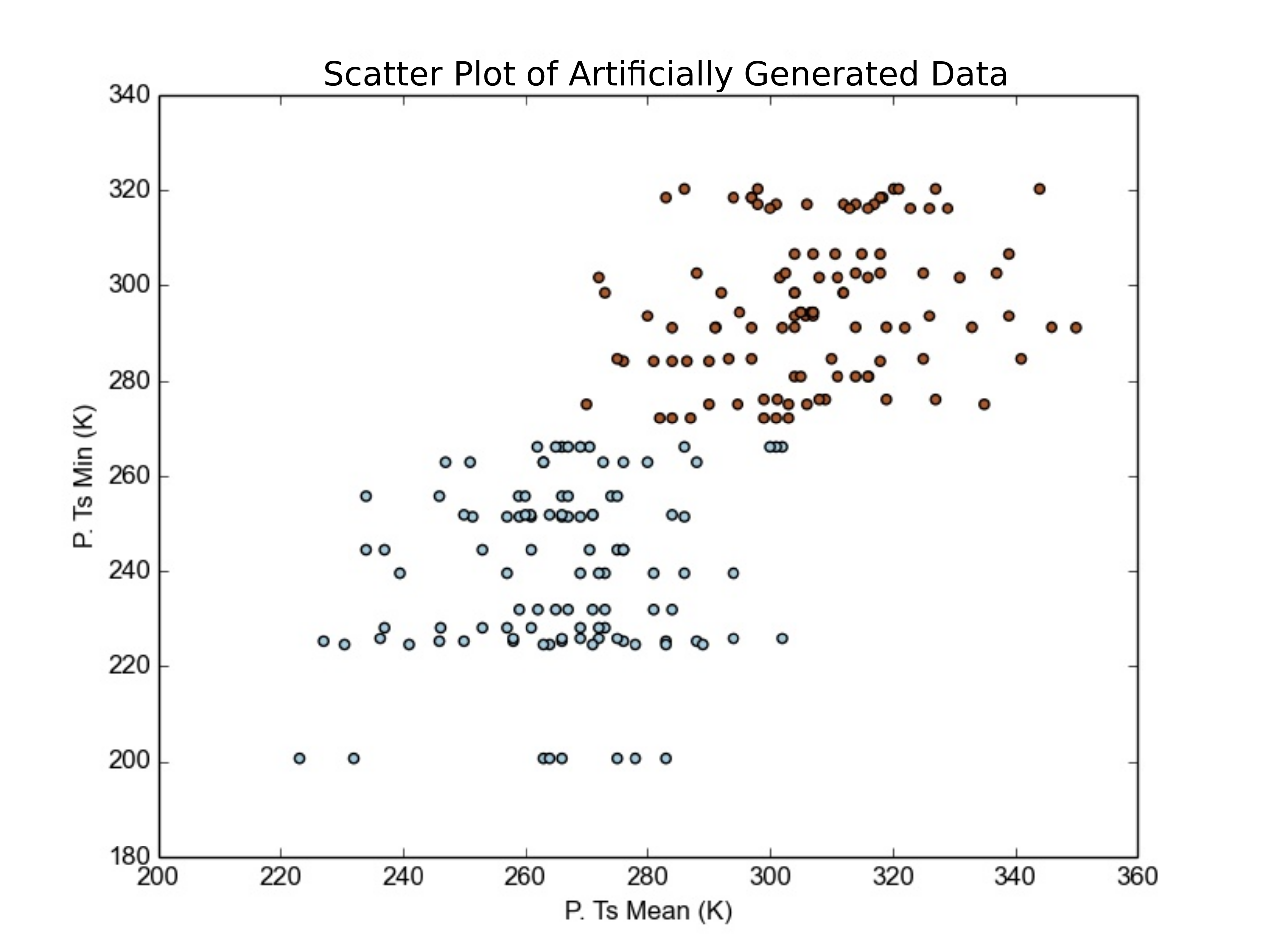}
	\caption{Scatter plot of newly generated artificial data points in two dimensions.}
\end{subfigure}
~
\begin{subfigure}[t]{0.45\columnwidth}
	\includegraphics[width=\columnwidth]{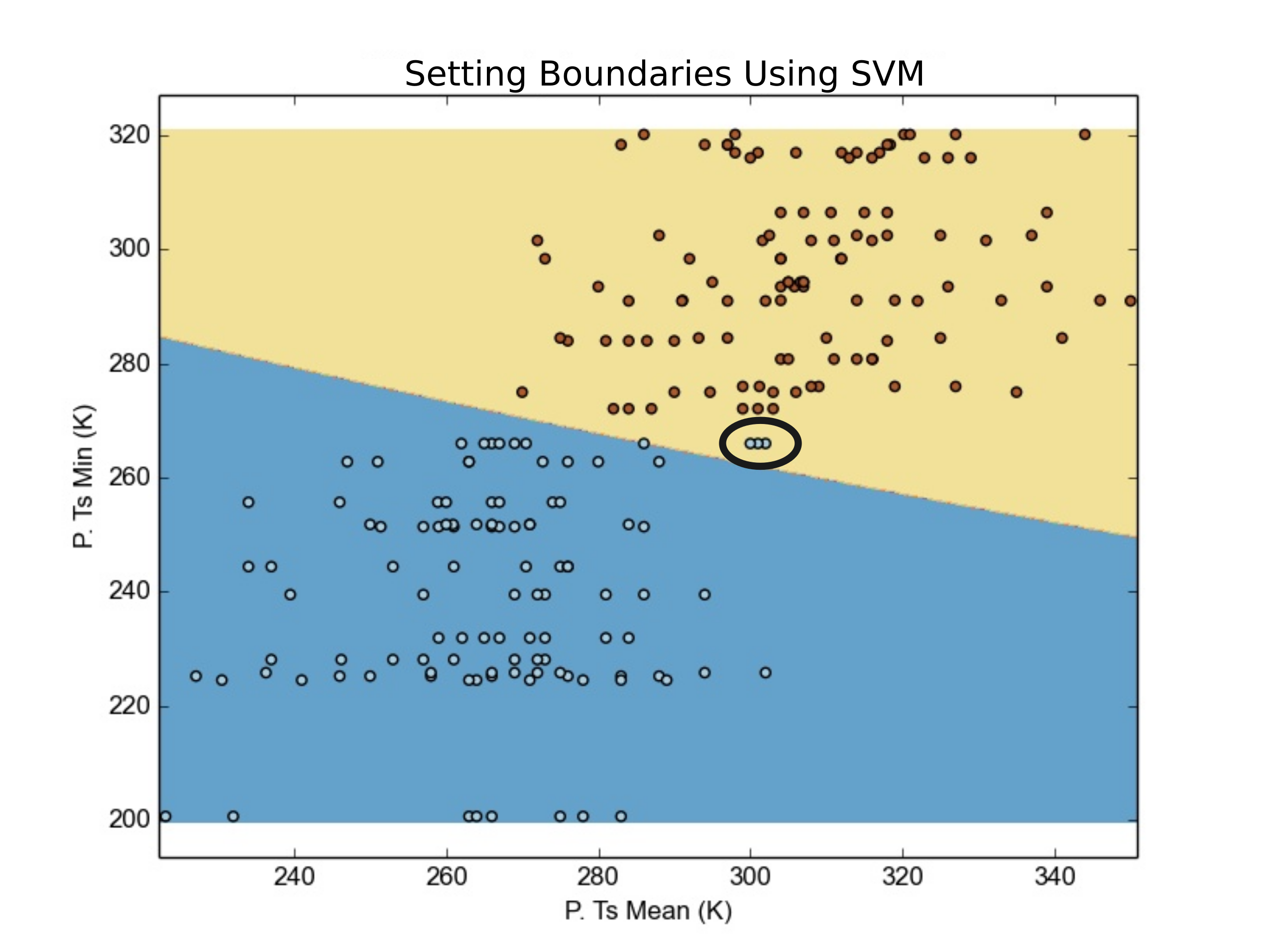}
    \caption{The best boundaries between two classes are determined using SVM. Here, there are three non-conforming data points (encircled) belonging to the mesoplanet class.}
\end{subfigure}

\begin{subfigure}[t]{0.45\columnwidth}
	\includegraphics[width=\columnwidth]{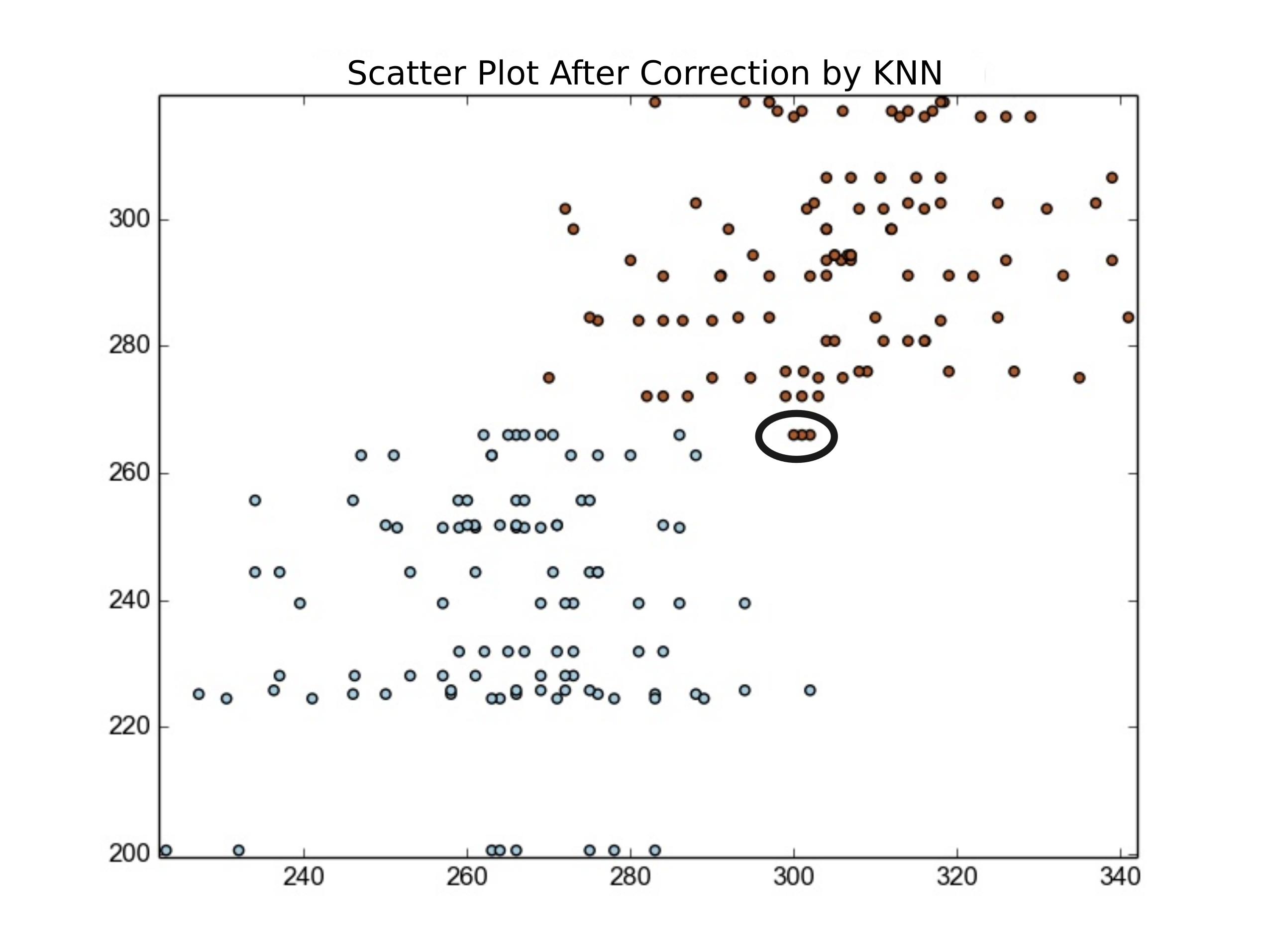}
	\caption{The three non-conforming data points' class belongingness rectified using K-NN. Now they belong to the class of psychroplanets, as their properties better reflect those of psychroplanets.}
\end{subfigure}
~
\begin{subfigure}[t]{0.45\columnwidth}
	\includegraphics[width=\columnwidth]{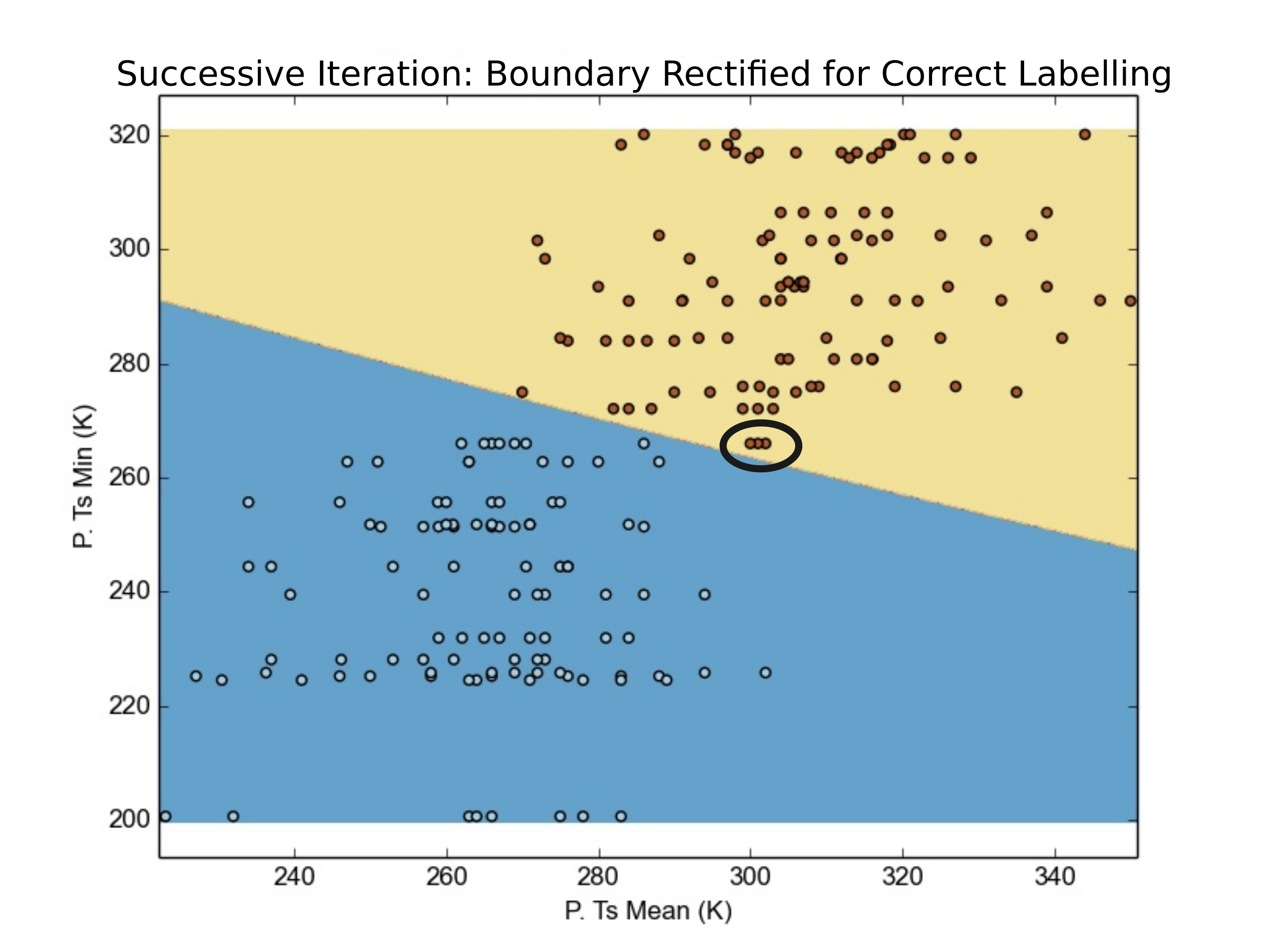}
	\caption{In the successive iteration, the boundary between the two classes has been adjusted to accommodate the three rectified points better. Now it is evident that the regions of the two classes (blue for mesoplanets and yellow for psychroplanets) comprise wholly of points that reflect the properties of the classes they truly belong to.}
\end{subfigure}

\caption{A new set of artificial data points being processed and their class-belongingness corrected in successive iterations of the SVM-KNN hybrid algorithm as a method of bounding and ensuring the purity of synthetic data samples.}
\label{fig:aug_method_steps}
\end{center}
\end{figure*}

The process of artificially generating and labeling data is illustrated in Figure \ref{fig:aug_method_steps}. A new set of data points generated randomly from the estimated Poisson distributions of both classes are plotted. The points in red depict artificial points belonging to the class of psychroplanets and the points in blue depict artificial points belonging to the class of mesoplanets. In general, the number of non-conforming points are less in number as the estimated distributions of the features of either class are different. Figure \ref{fig:aug_method_steps}(b) depicts three points (encircled) that should belong to the psychroplanet class but belongs to the mesoplanet class: note that these three points cross the boundary between the two classes as set by an SVM. The blue portion may contain points which belong to only the mesoplanet class and the yellow portion may contain points which belong only to the psychroplanet class, but these three points are non-conforming according to the boundary imposed. Hence, in order to ascertain the correct labels, these three points are subjected to a K-NN based rectification. In Figure \ref{fig:aug_method_steps}(c), the points in the dataset are plotted after being subjected to K-NN with $k = 5$ and class labels are rectified as is necessary to uphold the purity of either class. The three previously non-conforming points are determined to actually belong to the class of psychroplanets, and hence their class-belongingness is changed. Figure \ref{fig:aug_method_steps}(d) shows that the boundary between the two classes is altered by incorporating the rectified class-belongingness of the previously non-conforming points. In this figure, it is to be noted that all the points are conforming, and there are no points which belong to the region of the wrong class.

\begin{figure}[htbp!]
\begin{center}
\includegraphics[width=0.7\columnwidth]{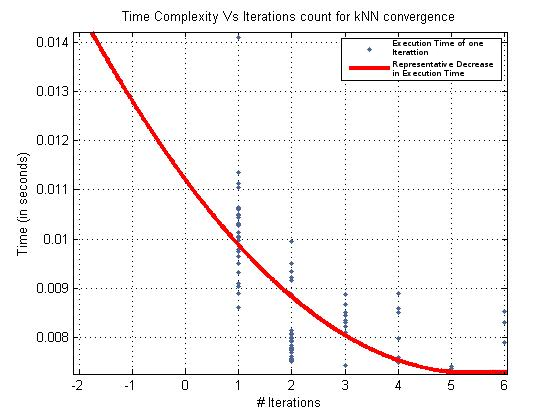}
\caption{A quadratic curve has been fit to the execution times of successive iterations in a run of the SVM-KNN method to demonstrate the general trend of decreasing times of execution of successive iterations. The time required to converge to the perfect labeling of class-belongingness of the synthetic data points reduces with each successive iteration resulting in the dip exhibited in successive iterations. This fortifies the efficiency of the proposed hybrid SVM-KNN algorithm.}
\label{fig:lcfit}
\end{center}
\end{figure}

This procedure was run many times on the artificially generated data to estimate the number of iterations and the time required for each iteration until the resulting dataset was devoid of any non-conforming data points. As the process is inherently stochastic, each new run of the SVM-KNN algorithm might result in a different number of iterations (and different amounts of execution time for each iteration) required until zero non-conforming samples are achieved. However, a general trend may be analyzed for the purpose of ascertaining that the algorithm will complete in a finite amount of time. Figure \ref{fig:lcfit} is a plot of the $i^{th}$ iteration against the time required for the algorithm to execute the respective iteration (to rectify the points in the synthesized dataset). From this figure, it should be noted that each successive iteration requires a smaller amount of time to complete: the red curve (a quadratic fit of the points) represents a decline in the time required for the SVM-KNN method to complete execution in successive iterations of a run. The number of iterations required for the complete execution of the SVM-KNN method ranges from one to six, with a generally declining execution time of successive iterations, proving the stability of the hybrid algorithm. Any algorithm is required to \textit{converge}: a point beyond which the execution of the algorithm ceases. In this case, convergence must ensure that every artificially generated data point conforms to the general properties of the class to which it is labeled to belong.

\subsection{Generating Data Empirically}

\label{sec:appendix_empirical_distributions}

In Section \ref{sec:parzen_windows}, a technique to estimate the probability density function of a sequence of random variables, called KDE (using Parzen-window estimation), was briefly described. We elaborate on the method and validate the method by performing a density estimation on random numbers generated from several known analytic continuous and discrete distributions. As a part of this exercise, we have presented goodness-of-fit scores for the estimated distributions. In addition to that, we have also plotted the graphs of the data points for a visualization of the density (generated using the standard analytic distributions as well as a Parzen-window estimate of those distributions).

We provide evidence of the working of Parzen-window estimation for different standard continuous and discrete probability distributions.

\subsubsection{An Analytical Exposition of Density Exploration}

Let $X = x_1, x_2, \dots, x_n$ be a sequence of independent and identically distributed multivariate random variables having $d$ dimensions. The window function used is a variation of the uniform kernel defined on the set $R^d$ as follows:
\begin{equation} \label{eq:parz_ker}
	\phi(u) =
    \begin{cases}
		1 & \quad u_j \leq \frac{1}{2} \quad \forall j \in \{1, 2, \dots, d\} \\
        0 & \quad otherwise
	\end{cases}
\end{equation}
Additionally, another parameter, the edge length vector $h = \{h_1, h_2, \dots h_d\}$, is defined, where each component of $h$ is set on a heuristic that considers the values of the corresponding feature in the original data. If $f_j$ is the column vector representing some feature $j \in X$ and
\begin{equation}
\begin{split}
	l_j &= min\{{(a-b)}^2 \quad \forall\ a,b \in f_j\}\\
	u_j &= max\{{(a-b)}^2 \quad \forall\ a,b \in f_j\},
\end{split}
\end{equation}
the edge length $h_j$ is given by,
\begin{equation}
	h_j = c \left(\frac{u_j + 2l_j}{3}\right)
\end{equation}
where $c$ is a scale factor.

Let $x' \in R^d$ be a random variable at which the density needs to be estimated. For the estimate, another vector $u$ is generated whose elements are given by:
\begin{equation}
	u_j = \frac{{x_j}' - x_{ij}}{h_j} \qquad \forall j \in \{1, 2, \dots, d\}
\end{equation}
The density estimate is then given by the following relationship:
\begin{equation} \label{eq:parz_main}
	p(x') = \frac{1}{n\prod_{i=1}^{d}h_i}\sum^{n}_{i=1}\phi(u)
\end{equation}

\subsubsection{Generating Synthetic Samples From the Estimated Empirical Distribution}

Traditionally, random numbers are generated from an analytic density function by inversion sampling. However, this would not work on a numeric density function unless the quantile function is numerically approximated by the density function. In order to avoid this, a form of rejection sampling has been used.

Let $r$ be a $d$-dimensional random vector with each component drawn from a uniform distribution between the minimum and maximum value of that component in the original data. Once the density, $p(r)$ is estimated by Equation \eqref{eq:parz_main}, the probability is approximated to:
\begin{equation} \label{eq:parz_pr}
	Pr(r) = p(r)\prod_{j=1}^{d}h_j
\end{equation}

To either accept or reject the sample $r$, another random number is generated from a uniform distribution within the range $[0,1)$. If this number is greater than the probability estimated by Equation \eqref{eq:parz_pr}, then the sample is accepted. Otherwise, it is rejected.

For the PHL-EC dataset, synthetic data was generated for the mesoplanet and psychroplanet classes using this method by taking $c=4$ for mesoplanets and $c=3$ for psychroplanets (in Equation \eqref{eq:parz_main}). 1000 samples were then generated for each class using rejection sampling on the density estimate. In this method, the bounding mechanism was not used and the samples were drawn out of the estimated density. Here, the top 85$\%$ of the features by importance were considered to estimate the probability density, and hence the boundary between the two classes using SVM was not constructed. The values of the remaining features were copied from the naturally occurring data points and shuffled between the artificially augmented data points in the same way as in the method described in Section \ref{subsec:data_aug-svmknn}). The advantage of using this method is that it may be used to estimate a distribution which resembles more closely the actual distribution of the data. However, this process is a little more complex than assuming a standard probability distribution in the data. Nonetheless, this is an inherently unassuming method and can accommodate distributions in data which are otherwise difficult to describe using the commonly used methods for describing the density of data.

\subsubsection{Parzen-Window Estimation of Continuous Random Variables}
For the tests involving univariate continuous random variables, standard distributions were used with the location parameter set to 0 and the scale set to 1. The distributions used in their univariate forms were:
\begin{enumerate}
	\item Normal
	\item Cauchy
	\item Laplace
	\item Rayleigh
\end{enumerate}

For multivariate continuous random variables, however, the test was performed on the normal density for which the mean and covariance were explicitly set to the following values:
\begin{equation}
\begin{split}
	\mu &= \{0.5, -0.2\}\\
	Z &=
	\begin{bmatrix}
	2.0 & 0.3\\
	0.3 & 0.5
	\end{bmatrix}
\end{split}
\end{equation}

\begin{figure*}
\centering
\begin{subfigure}{0.45\columnwidth}
	\includegraphics[width=\textwidth]{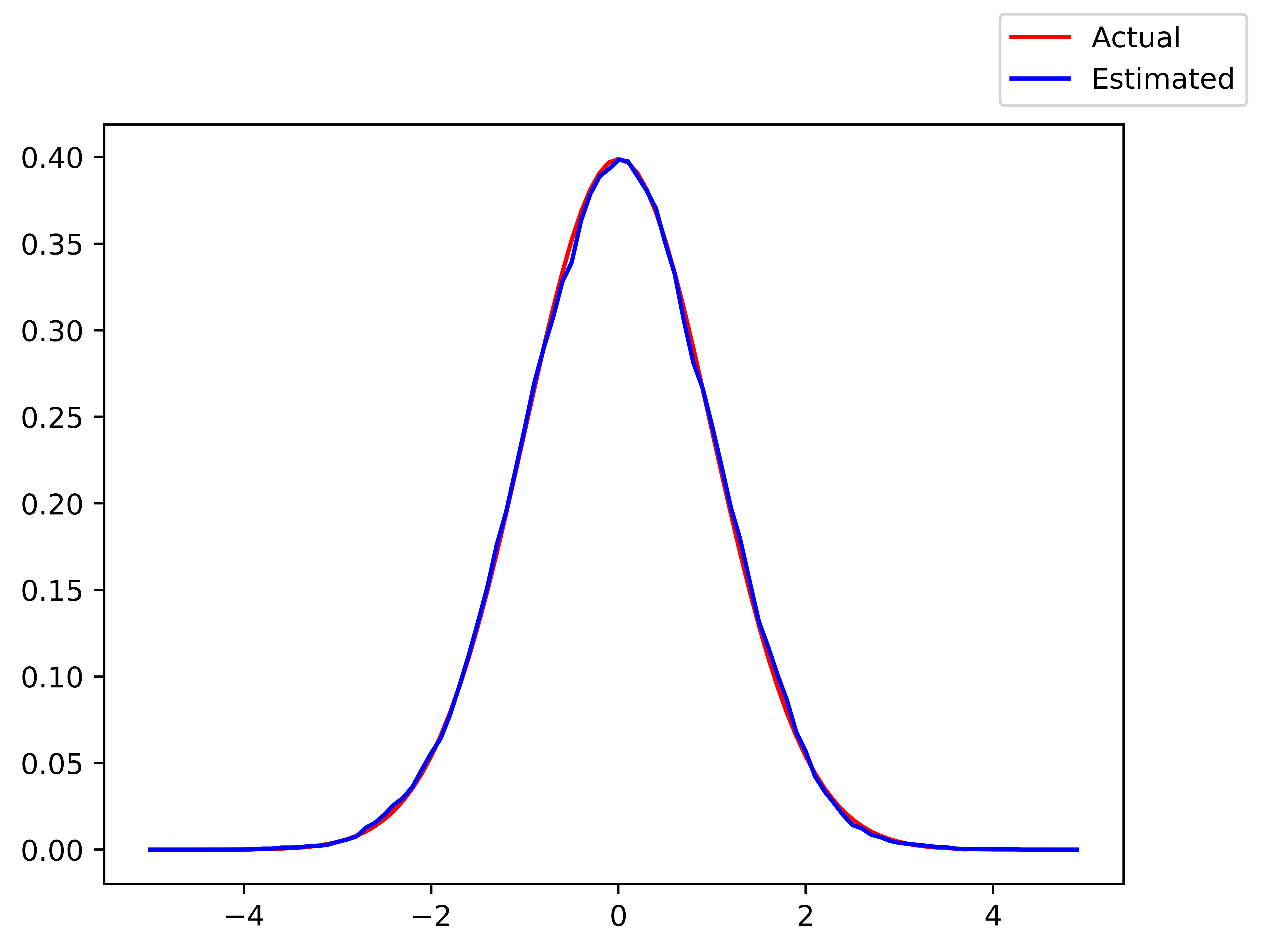}
	\caption{Normal}
\end{subfigure}
~~
\begin{subfigure}{0.45\columnwidth}
	\includegraphics[width=\textwidth]{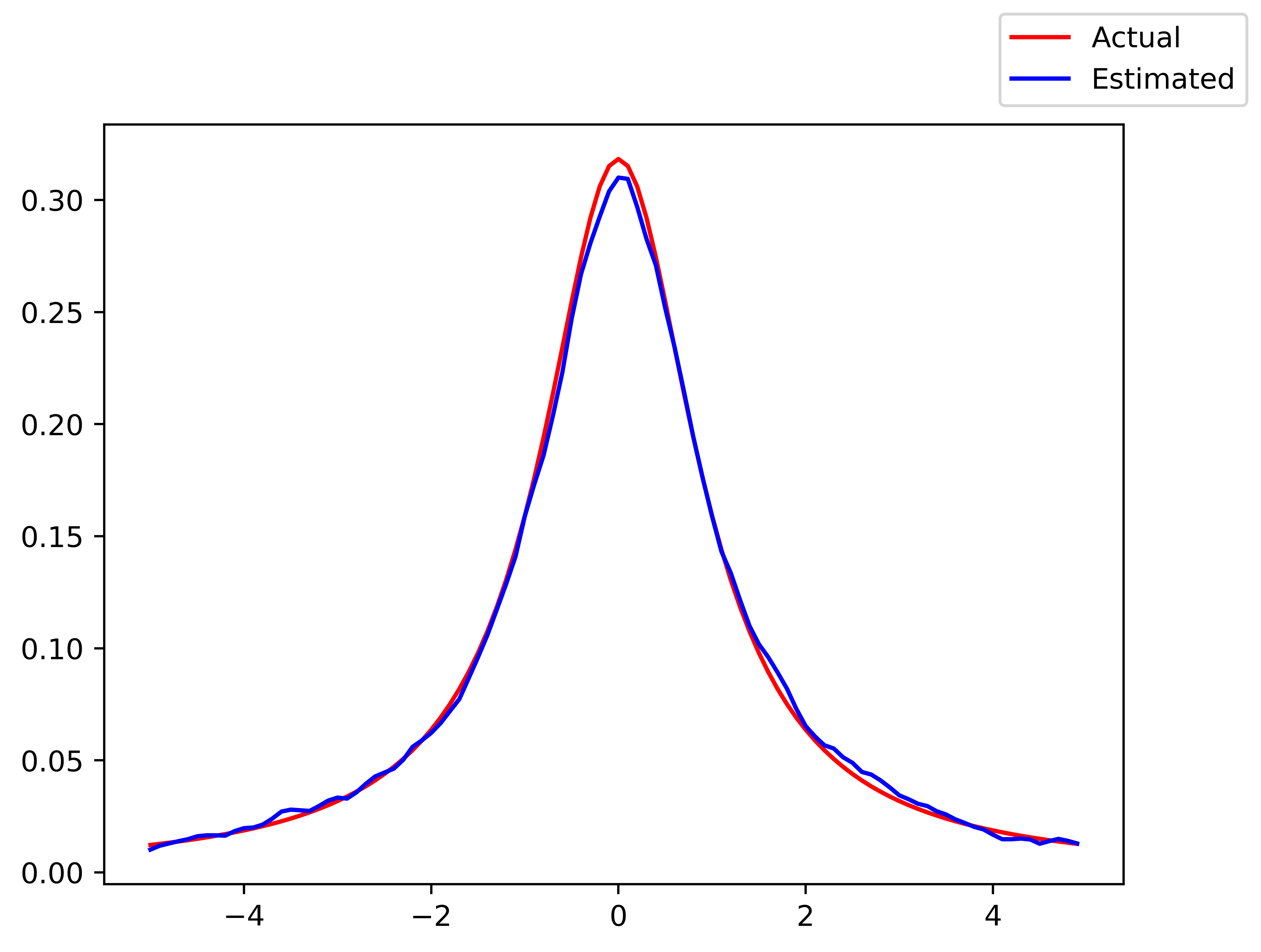}
    \caption{Cauchy}
\end{subfigure}

\begin{subfigure}{0.45\columnwidth}
	\includegraphics[width=\textwidth]{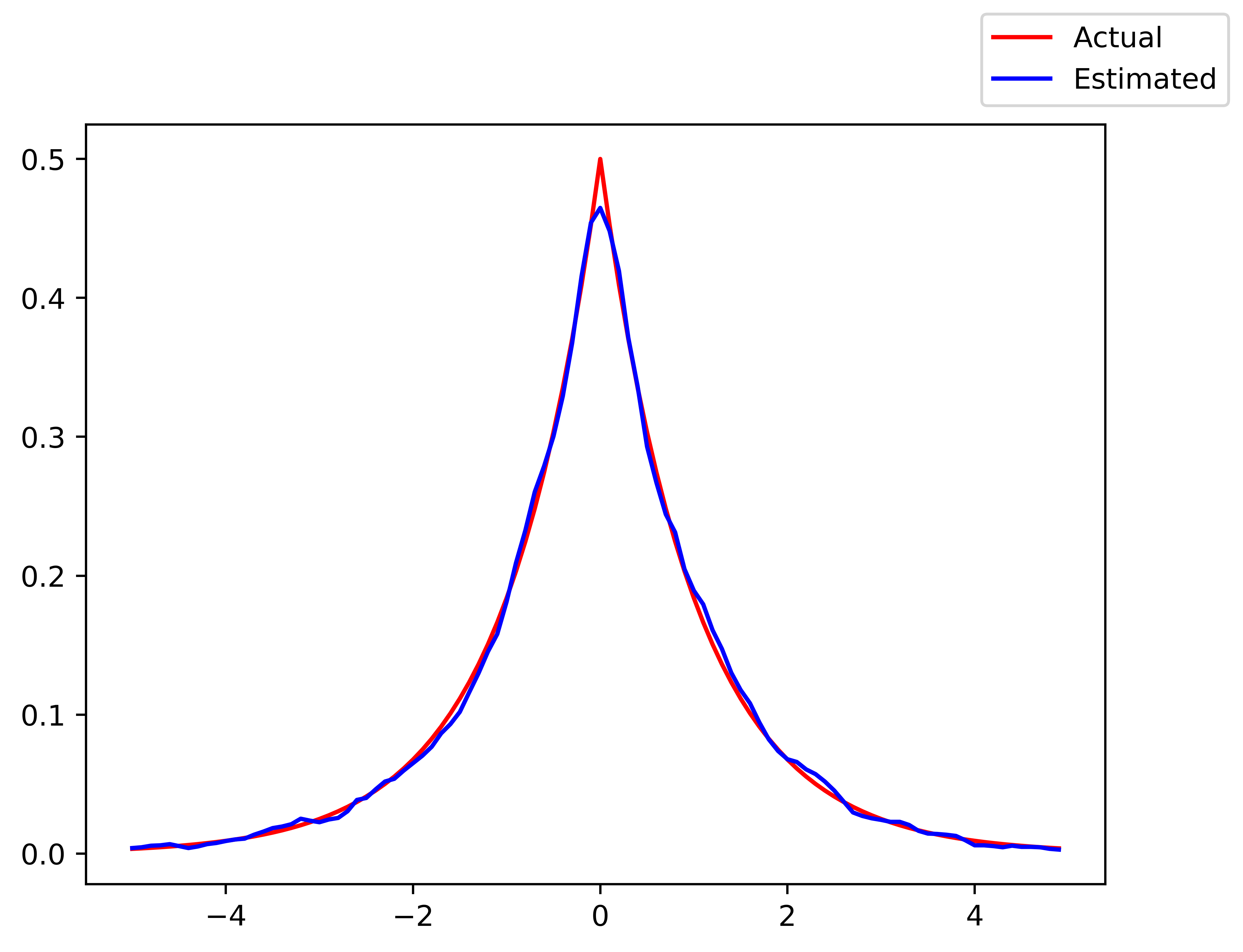}
	\caption{Laplace}
\end{subfigure}
~~
\begin{subfigure}{0.45\columnwidth}
	\includegraphics[width=\textwidth]{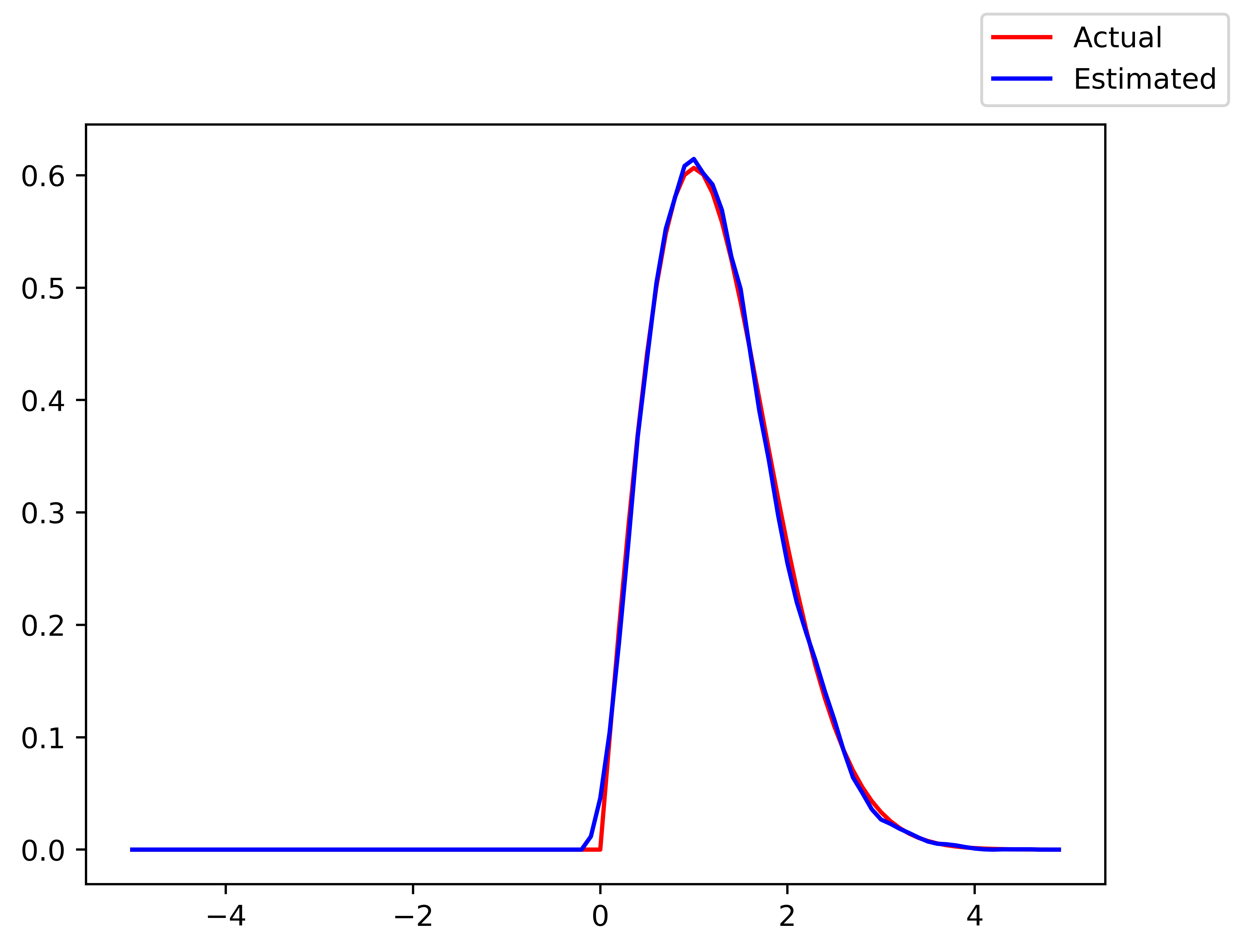}
	\caption{Rayleigh}
\end{subfigure}
\caption{Estimates for univariate continuous distributions}
\label{fig:ucont}
\end{figure*}

\begin{figure*}
\centering
\begin{subfigure}{0.45\columnwidth}
	\includegraphics[width=\textwidth]{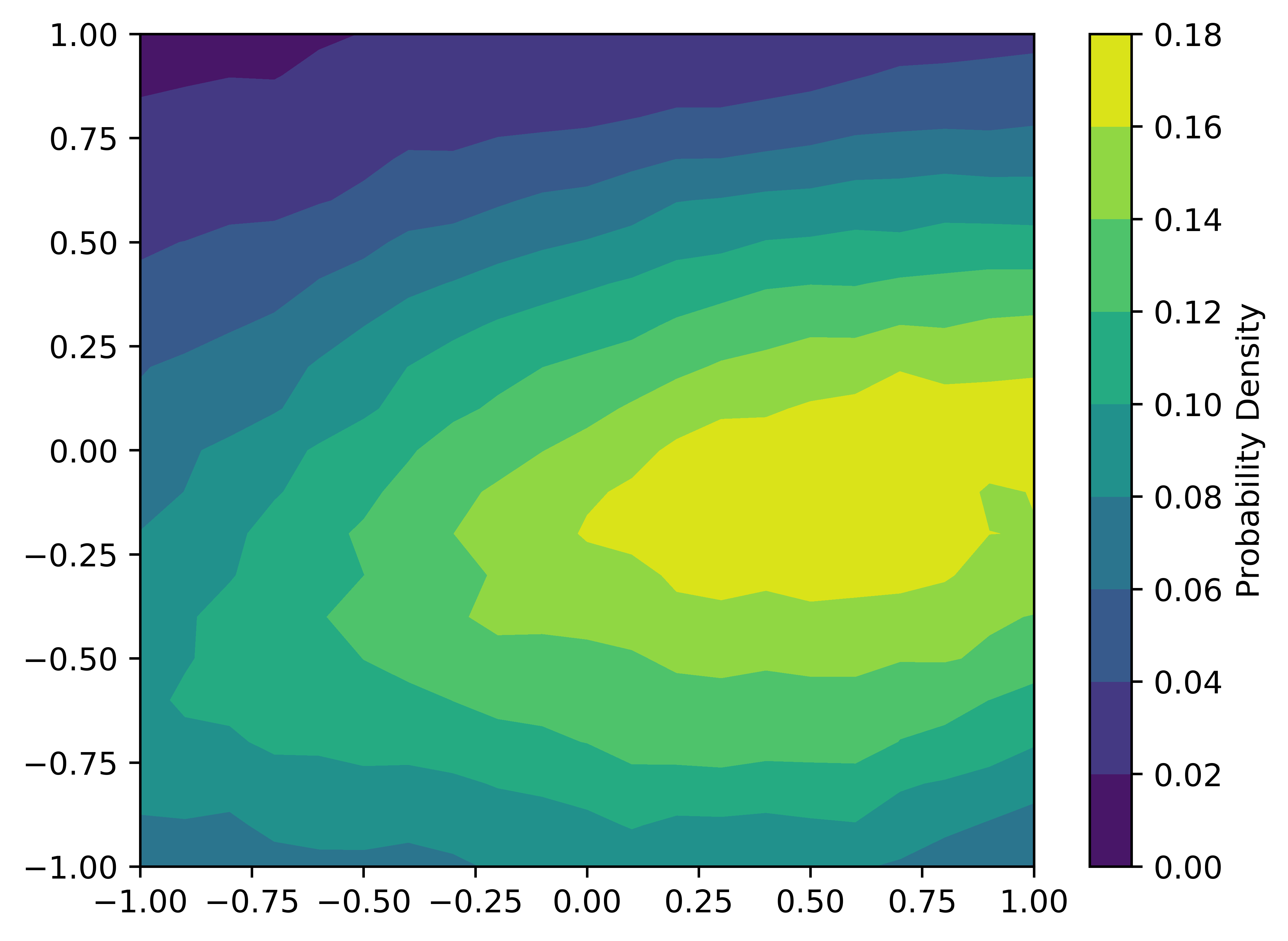}
	\caption{Estimated}
\end{subfigure}
~
\begin{subfigure}{0.45\columnwidth}
	\includegraphics[width=\textwidth]{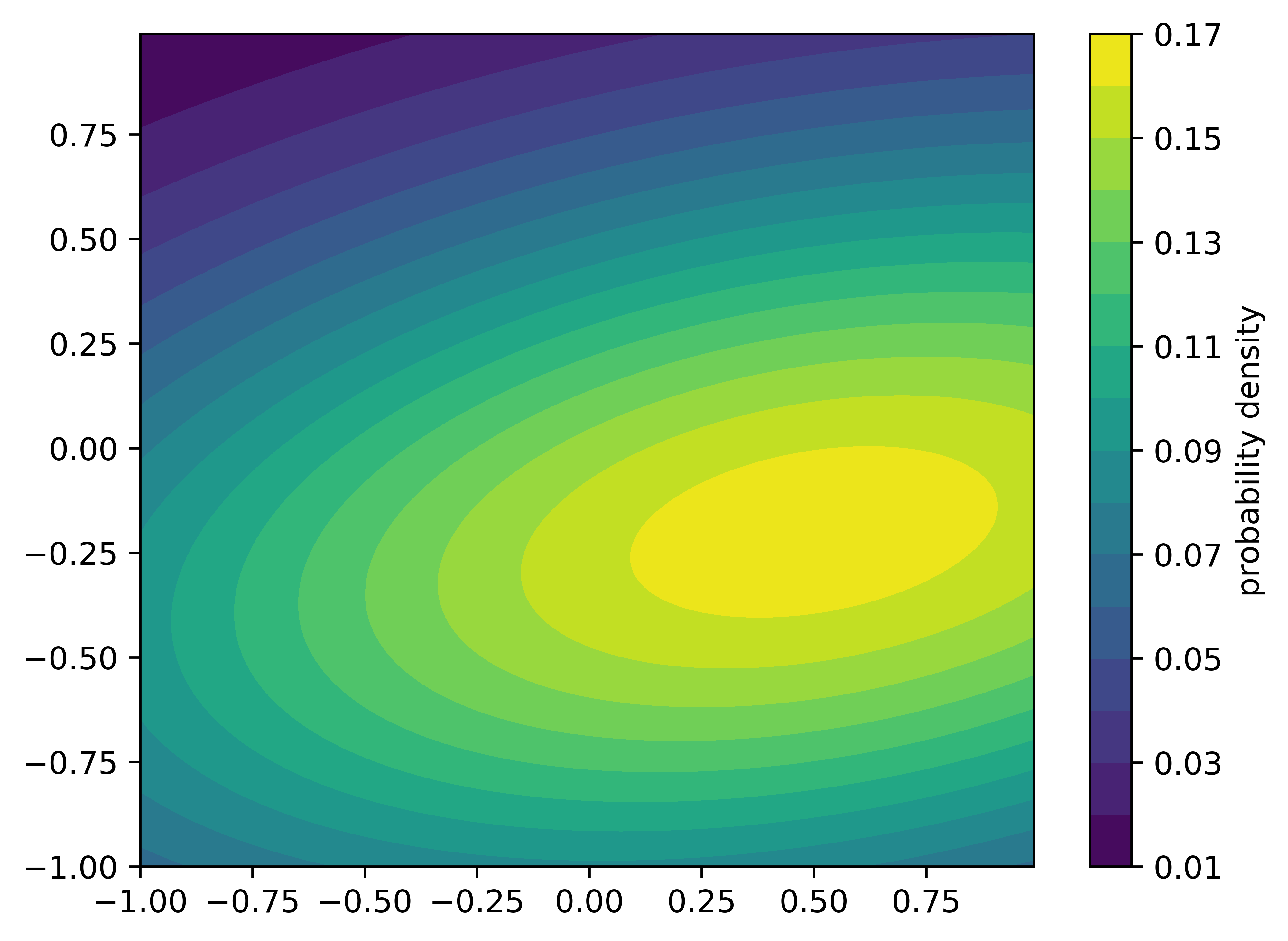}
	\caption{Actual}
\end{subfigure}
\caption{Multivariate normal density estimation in two dimensions.}
\label{fig:mult}
\end{figure*}

For each distribution, 10000 samples were generated and then used to estimate the original analytic distribution function. For the univariate case, the actual and estimated probability density were calculated over the range -5 to 5 by stepping at 0.1 and for multivariate, the same was done over the range (-1, -1) to (1, 1) by stepping at 0.1. The mean squared error was calculated and has been presented in Table \ref{t:mse}. The graphs of these calculated values are presented in Figure \ref{fig:ucont} for univariate data and Figure \ref{fig:mult} for multivariate data.

\begin{table}
\centering
\begin{tabular}{| c c c |}
	\hline
	Distribution & $c$ value & MSE ($\times 10^-5$)\\
	\hline
    Normal (univariate) & 0.2 & 0.8164\\
    Normal (multivariate) & 0.23 & 3.3946\\
	Cauchy & 0.0002 & 1.7011\\
	Laplace & 0.07 & 3.3099\\
	Rayleigh & 0.15 & 4.3095\\
	\hline
\end{tabular}
\caption{Mean squared error for continuous standard distributions.}
\label{t:mse}
\end{table}

\begin{figure*}
\centering
\begin{subfigure}{0.45\columnwidth}
	\includegraphics[width=\textwidth]{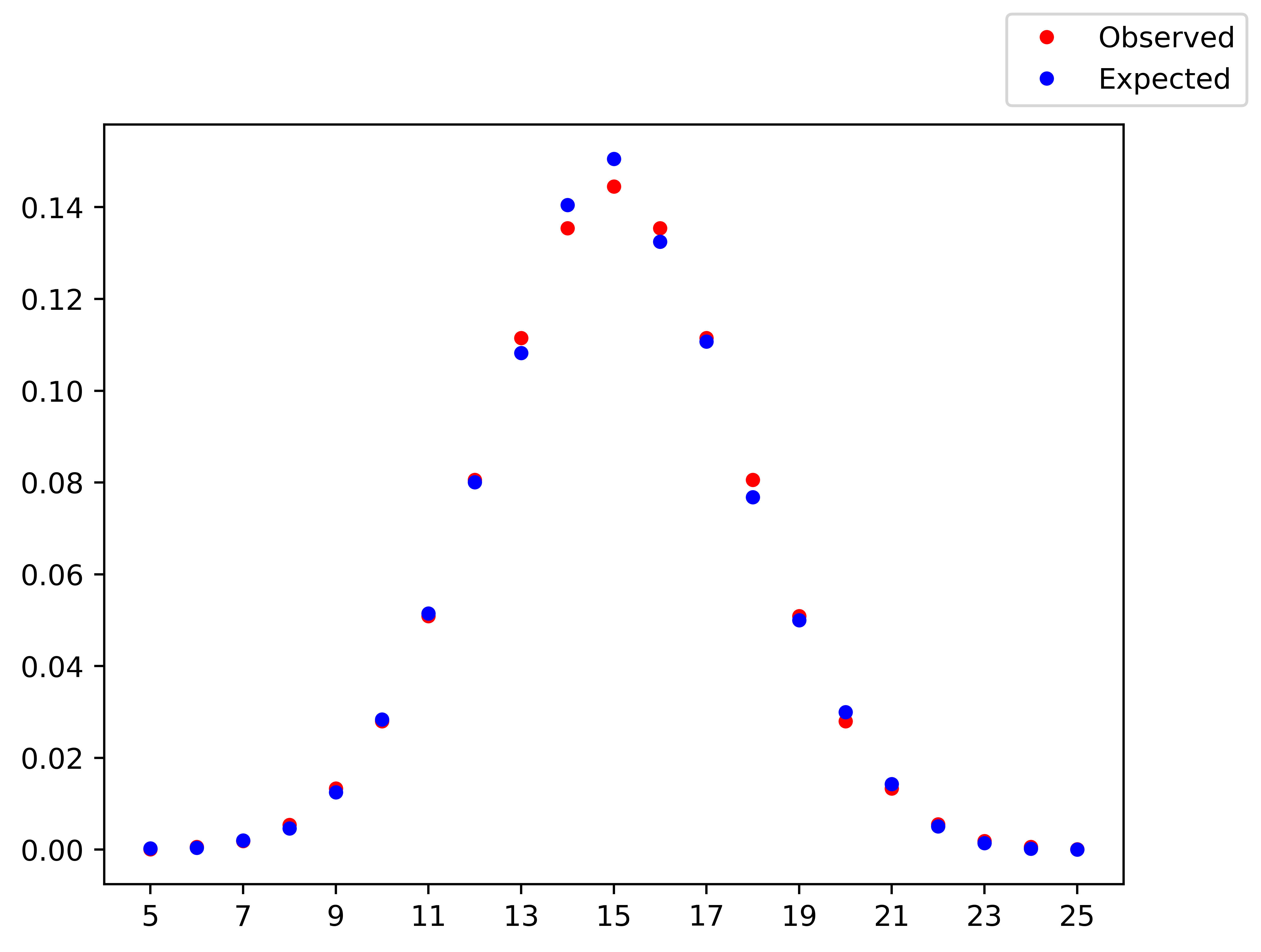}
	\caption{Binomial}
\end{subfigure}
~~
\begin{subfigure}{0.45\columnwidth}
	\includegraphics[width=\textwidth]{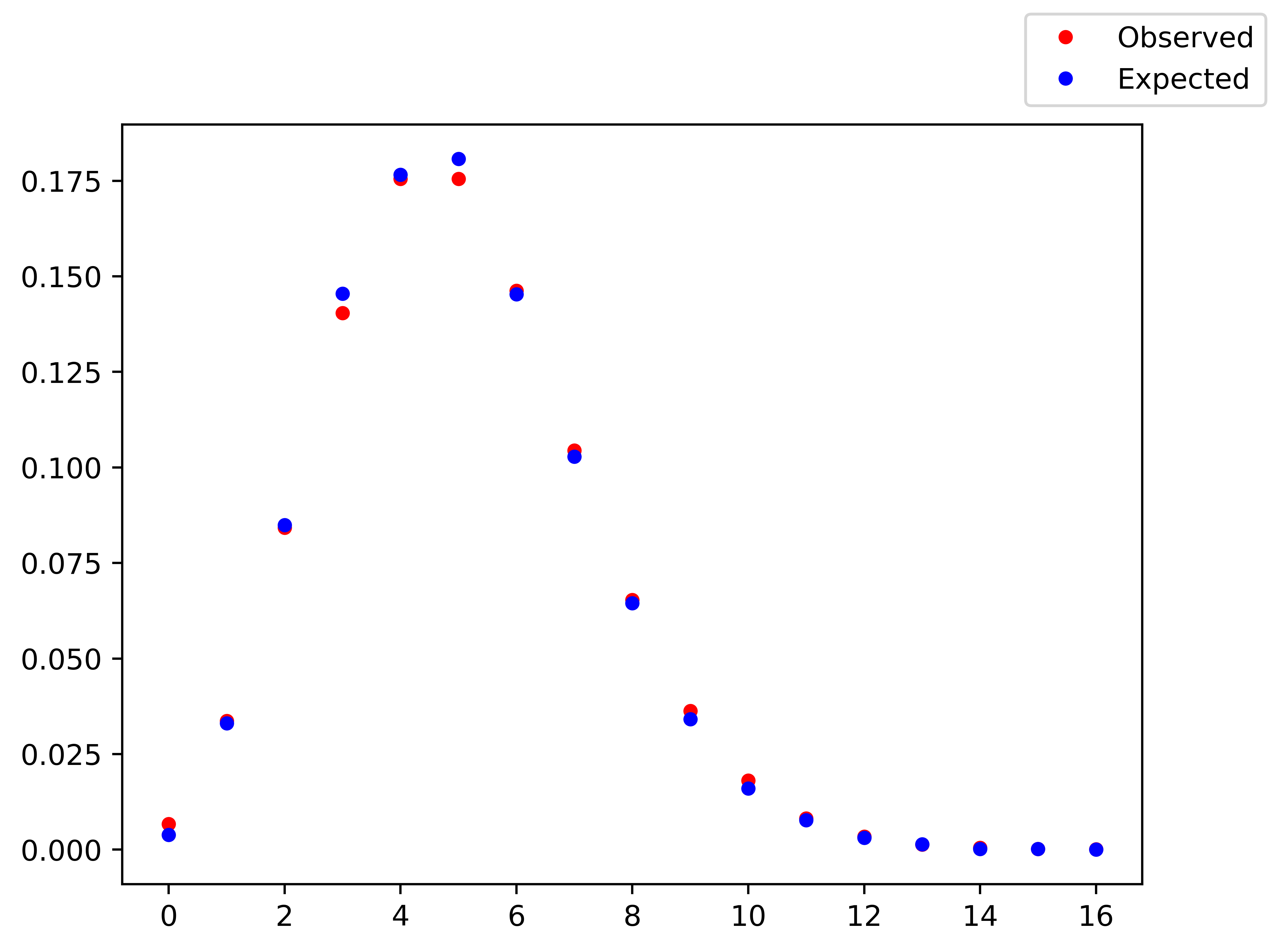}
	\caption{Poisson}
\end{subfigure}
\caption{Estimates for discrete distributions}
\label{fig:disc}
\end{figure*}

\subsubsection{Parzen-Window Estimation of Discrete Random Variables}
The discrete probability distributions used for testing KDE and their parameters were:
\begin{enumerate}
	\item Binomial (n=30, p=0.5)
	\item Poisson ($\mu$=5)
\end{enumerate}

These distributions were tested by generating 10000 samples each, which were then used to estimate the mass function of the original distribution. Once the mass function was estimated, 10000 random samples were drawn from the estimated distribution using rejection sampling. Further, for the binomial distribution, the probabilities were estimated for integers from 5 to 25, both inclusive, while for the Poisson distribution, the probabilities were estimated for integers in the range 0 to 16, both inclusive. Once these expected and observed values were generated, a chi-squared test was performed and the results have been laid out in Table \ref{t:chi}. The frequency graphs for both observed and expected values are presented in Figure \ref{fig:disc}.

\begin{table}
\centering
\begin{tabular}{| c c c |}
	\hline
    Distribution & $c$ value & $\chi^2$\\
    \hline
    Poisson & $0.05$ & $22.94$\\
	Binomial & $0.01$ & $22.02$\\
    \hline
\end{tabular}
\caption{Chi Square test results for discrete distributions.}
\label{t:chi}
\end{table}

In the cases of both discrete and continuous distributions, it is evident that KDE's performance is reasonable for generating data points. A small MSE or Chi-squared score is desirable as this represents a small deviation of artificially generated points from the actual distribution of the data. Thus, in experiments which require the estimation of the distribution of data, KDE is a candidate whose efficacy may be compared with the common methods of assuming a probability density using well established probability densities.

%\section{Salient Features of the Thermal Suitability Index}

\section{Future Work: Activation Function (SBAF) for a Neural Network} \label{app:sbaf}
The investigation into finding an optimal set of features for surface temperature based discrimination of habitability (Section \ref{subsec:tss}) led us to modeling an activation function which may be used in back propagation in Neural Networks. The activation function may be used to reduce the error in back propagation while using Artificial Neural Networks to classify exoplanets based on habitability features. The structure of the activation is inspired by the outcome of TSS (Table \ref{tab:tss}) and previous work on habitability modeling and classification \citep{CDHPF2016,Proxb} based on econometric production function \citep{Saha2016,Ginde2016,ginde2015mining,1610.00624}. The visualization of the activation function shows that it doesn't suffer from the local oscillation problem and may not experience premature convergence witnessed in gradient descent/ascent based methods. We present a nice analytical property of the function below.
\begin{equation}
\begin{split}
y &= \frac{1}{1 + kx^{\alpha}(1-x)^{1-\alpha}}\\
\Rightarrow \textrm{ln}y &= \textrm{ln}y - \textrm{ln}(1 + kx^{\alpha}(1-x)^{1-\alpha})\\
 &= - \textrm{ln}(1 + kx^{\alpha}(1-x)^{1-\alpha})\\
\Rightarrow \frac{1}{y}\frac{dy}{dx} &= - \frac{1}{(1 + kx^{\alpha}(1-x)^{1-\alpha})} \cdot \Big[k\alpha x^{\alpha -1} (1-\alpha)^{1-\alpha} - kx^{\alpha}(1-\alpha)(1-x)^{1 - \alpha -1} \Big]\\
&= - \frac{k}{(1 + kx^{\alpha}(1-x)^{1-\alpha})} \cdot \Big[\alpha x^{\alpha -1} (1-\alpha)^{1-\alpha} - (1-\alpha)x^{\alpha}(1-x)^{-\alpha} \Big]\\
\Rightarrow \frac{dy}{dx} &= y \Bigg[ \frac{\alpha}{x} - (1 - \alpha)\frac{1}{1-x} \Bigg]kx^{\alpha}(1-x)^{1-\alpha}\\
&= y \Bigg[\frac{\alpha (1-x) - (1-\alpha)x}{x(1-x)} \Bigg]kx^{\alpha}(1-x)^{1-\alpha}\\
&= y^{2} \Bigg[\frac{\alpha - x}{x(1-x)} \Bigg]kx^{\alpha}(1-x)^{1-\alpha}\\
\end{split}
\label{eq:acti_func_1}
\end{equation}

From the definition of the function, we have:
\begin{equation}
\begin{split}
y &= \frac{1}{1 + kx^{\alpha}(1-x)^{1-\alpha}}\\
\Rightarrow kx^{\alpha}(1-x)^{1-\alpha} &= \frac{1-y}{y}
\end{split}
\label{eq:acti_func_2}
\end{equation}

Substituting Equation \ref{eq:acti_func_2} in \ref{eq:acti_func_1},

\begin{equation}
\begin{split}
\frac{dy}{dx} &= y^{2} \cdot \frac{\alpha-x}{x(1-x)} \cdot \frac{1-y}{y}\\
&= \frac{y(1-y)}{x(1-x)}\cdot(\alpha-x)
\end{split}
\label{eq:acti_func_first_deriv}
\end{equation}

From Equation \ref{eq:acti_func_first_deriv}, the second derivative is computed as follows:

\begin{equation}
\begin{split}
\frac{d^{2}y}{dx^{2}} = \frac{y(y-1) \big[x(1-x) + (\alpha - x)(1-2x) - (\alpha - x)^{2}(1-2y)\big]}{x^{2}(1-x)^{2}}
&= \frac{y(y-1)}{x(1-x)}; (\alpha=x)
\end{split}
\label{eq:acti_func_second_deriv}
\end{equation}
% \frac{d^{2}y}{dx^{2}} = \frac{y(y-1) \big[x(1-x) + (\alpha - x)(1-2x) - (\alpha - x)^{2}(1-2y)\big]}{x^{2}(1-x)^{2}}
% \frac{d^{2}y}{dx^{2}} = \frac{y(y-1) \big[x(1-x) + (\alpha - x)(1-2x) - (\alpha - x)^{2}(1-2y)\big]}{x^{2}(1-x)^{2}}
Clearly, the first derivative vanishes when $\alpha=x$, the derivative is positive when $\alpha > x$ and is negative when $\alpha < x$ (implying range of values for $\alpha$ so that the function becomes increasing or decreasing). We need to determine the sign of the second derivative when $\alpha=x$ to ascertain the condition of maxima (corresponding to maximum width of the separating hyperplane ensuring optimal discrimination between habitability classes). Assuming $ 0< x <1$, the condition of optimality, $0 \leq \alpha \leq 1$, $y$ by construction lies between $(0,1)$. Hence, $\frac{d^{2}y}{dx^{2}} < 0$ ensuring maxima of $y$.
\begin{itemize}
\item $x$ is surface temperature (normalized between $0$ and $1$) and $1-x$ is the complement of that, together explaining the perfect discrimination between habitability classes as explained in our TSS above. The motivation of SBAF is derived from this fact of TSS. Using $kx^{\alpha}(1-x)^{1-\alpha}$ shall maximize the width of the two separating hyperplanes in the SVM used in TSS (See the proof below) as the kernel has a global maxima when $0 \leq \alpha \leq 1$. This is equivalent to the CDHS formulation when CD-HPF is written as $ y = kx^{\alpha}(1-x)^{\beta}$ where $\alpha+\beta=1, 0 \leq \alpha \leq 1, 0 \leq \beta \leq 1$, $k$ is suitably assumed to be $1$ (CRS condition), and the representation ensures global maxima (maximum width of the separating hyperplanes) under such constraints \citep{CDHPF2016,Proxb}. 
\item The new activation function to be used for training a neural network for habitability classification boasts of an optima. Evidently, from the graphical simulations below, we observe less flattening of the function and therefore the formulation should be able to tackle local oscillations more easily as compared to the more generally used sigmoid function. Moreover, since $0 \leq \alpha \leq 1, 0 \leq x \leq 1, 0 \leq 1-x \leq 1$, the variable term in the denominator of SBAF, $kx^{\alpha}(1-x)^{1-\alpha}$ may be approximated to a first order polynomial. This may help us in circumventing expensive floating point operations without compromising the precision.
\item Need to show that the maxima is unique in the defined interval. This will circumvent the local maxima problem.
\end{itemize}
\textbf{:} 
\begin{figure}[htbp!]
\begin{center}
\includegraphics[width=0.8\columnwidth]{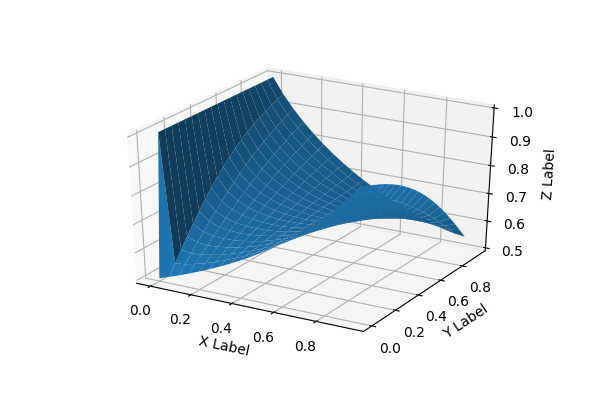}
\caption{application of SBAF on Mean Surface Temperature of exoplanets}
\label{fig:sbaf1}
\end{center}
\end{figure}

\begin{figure}[htbp!]
\begin{center}
\includegraphics[width=0.8\columnwidth]{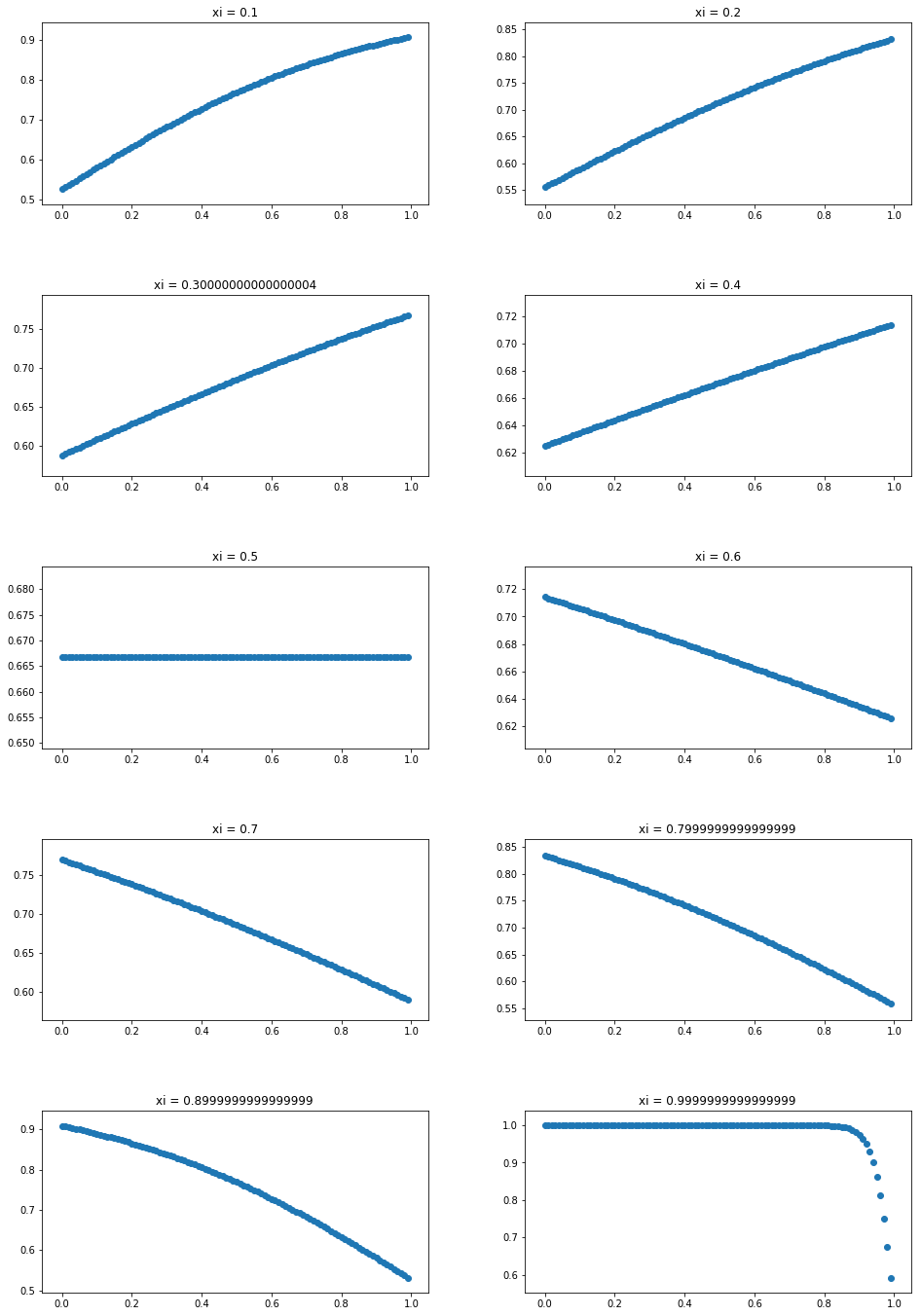}
\caption{}
\label{fig:sbaf2}
\end{center}
\end{figure}

\begin{comment}
\begin{figure}[htbp!]
\begin{center}
\includegraphics[width=0.5\columnwidth]{img2}
\caption{}
\label{fig:sbaf3}
\end{center}
\end{figure}
\end{comment}
\FloatBarrier

\section*{Bibliography} \label{sec:biblio}
\bibliography{mybibfile.bib}

\begin{thebibliography}{36}
\expandafter\ifx\csname natexlab\endcsname\relax\def\natexlab#1{#1}\fi
\providecommand{\url}[1]{\texttt{#1}}
\providecommand{\href}[2]{#2}
\providecommand{\path}[1]{#1}
\providecommand{\DOIprefix}{doi:}
\providecommand{\ArXivprefix}{arXiv:}
\providecommand{\URLprefix}{URL: }
\providecommand{\Pubmedprefix}{pmid:}
\providecommand{\doi}[1]{\href{http://dx.doi.org/#1}{\path{#1}}}
\providecommand{\Pubmed}[1]{\href{pmid:#1}{\path{#1}}}
\providecommand{\bibinfo}[2]{#2}
\ifx\xfnm\relax \def\xfnm[#1]{\unskip,\space#1}\fi
%Type = Article
\bibitem[{Bains and Schulze-Makuch(2016)}]{Bains2016}
\bibinfo{author}{Bains, W.}, \bibinfo{author}{Schulze-Makuch, D.},
  \bibinfo{year}{2016}.
\newblock \bibinfo{title}{The cosmic zoo: The (near) inevitability of the
  evolution of complex, macroscopic life}.
\newblock \bibinfo{journal}{Life} \bibinfo{volume}{6}, \bibinfo{pages}{25}.
\newblock \URLprefix \url{https://doi.org/10.3390/life6030025},
  \DOIprefix\doi{10.3390/life6030025}.
%Type = Article
\bibitem[{Barr et~al.(2017)Barr, Dobos and Kiss}]{Barr2017}
\bibinfo{author}{Barr, A.C.}, \bibinfo{author}{Dobos, V.},
  \bibinfo{author}{Kiss, L.L.}, \bibinfo{year}{2017}.
\newblock \bibinfo{title}{Interior structures and tidal heating in the
  {TRAPPIST}-1 planets}.
\newblock \bibinfo{journal}{Astronomy {\&} Astrophysics} \URLprefix
  \url{https://doi.org/10.1051/0004-6361/201731992},
  \DOIprefix\doi{10.1051/0004-6361/201731992}.
%Type = Article
\bibitem[{Bora et~al.(2016)Bora, Saha, Agrawal, Safonova, Routh and
  Narasimhamurthy}]{CDHPF2016}
\bibinfo{author}{Bora, K.}, \bibinfo{author}{Saha, S.},
  \bibinfo{author}{Agrawal, S.}, \bibinfo{author}{Safonova, M.},
  \bibinfo{author}{Routh, S.}, \bibinfo{author}{Narasimhamurthy, A.},
  \bibinfo{year}{2016}.
\newblock \bibinfo{title}{Cd-hpf: New habitability score via data analytic
  modeling}.
\newblock \bibinfo{journal}{Astronomy and Computing} \bibinfo{volume}{17},
  \bibinfo{pages}{129 -- 143}.
\newblock \URLprefix
  \url{http://www.sciencedirect.com/science/article/pii/S2213133716300865},
  \DOIprefix\doi{https://doi.org/10.1016/j.ascom.2016.08.001}.
%Type = Inproceedings
\bibitem[{Boser et~al.(1992)Boser, Guyon and Vapnik}]{Boser1992}
\bibinfo{author}{Boser, B.E.}, \bibinfo{author}{Guyon, I.M.},
  \bibinfo{author}{Vapnik, V.N.}, \bibinfo{year}{1992}.
\newblock \bibinfo{title}{A training algorithm for optimal margin classifiers},
  in: \bibinfo{booktitle}{Proceedings of the fifth annual workshop on
  Computational learning theory - {COLT} 1992}, \bibinfo{publisher}{{ACM}
  Press}.
\newblock \URLprefix \url{https://doi.org/10.1145/130385.130401},
  \DOIprefix\doi{10.1145/130385.130401}.
%Type = Article
\bibitem[{Breiman(1996)}]{Breiman1996}
\bibinfo{author}{Breiman, L.}, \bibinfo{year}{1996}.
\newblock \bibinfo{title}{Technical note: Some properties of splitting
  criteria}.
\newblock \bibinfo{journal}{Machine Learning} \bibinfo{volume}{24},
  \bibinfo{pages}{41--47}.
\newblock \URLprefix \url{https://doi.org/10.1023/A:1018094028462},
  \DOIprefix\doi{10.1023/A:1018094028462}.
%Type = Article
\bibitem[{Breiman(2001)}]{Breiman2001}
\bibinfo{author}{Breiman, L.}, \bibinfo{year}{2001}.
\newblock \bibinfo{title}{Random forests}.
\newblock \bibinfo{journal}{Machine Learning} \bibinfo{volume}{45},
  \bibinfo{pages}{5--32}.
\newblock \URLprefix \url{https://doi.org/10.1023/A:1010933404324},
  \DOIprefix\doi{10.1023/A:1010933404324}.
%Type = Article
\bibitem[{Chawla et~al.(2002)Chawla, Bowyer, Hall and
  Kegelmeyer}]{Chawla:2002:SSM:1622407.1622416}
\bibinfo{author}{Chawla, N.V.}, \bibinfo{author}{Bowyer, K.W.},
  \bibinfo{author}{Hall, L.O.}, \bibinfo{author}{Kegelmeyer, W.P.},
  \bibinfo{year}{2002}.
\newblock \bibinfo{title}{Smote: Synthetic minority over-sampling technique}.
\newblock \bibinfo{journal}{J. Artif. Int. Res.} \bibinfo{volume}{16},
  \bibinfo{pages}{321--357}.
\newblock \URLprefix \url{http://dl.acm.org/citation.cfm?id=1622407.1622416}.
%Type = Inproceedings
\bibitem[{Chen and Guestrin(2016)}]{Chen2016}
\bibinfo{author}{Chen, T.}, \bibinfo{author}{Guestrin, C.},
  \bibinfo{year}{2016}.
\newblock \bibinfo{title}{{XGBoost}}, in: \bibinfo{booktitle}{Proceedings of
  the 22nd {ACM} {SIGKDD} International Conference on Knowledge Discovery and
  Data Mining - {KDD} 2016}, \bibinfo{publisher}{{ACM} Press}.
\newblock \URLprefix \url{https://doi.org/10.1145/2939672.2939785},
  \DOIprefix\doi{10.1145/2939672.2939785}.
%Type = Article
\bibitem[{Cortes and Vapnik(1995)}]{Cortes1995}
\bibinfo{author}{Cortes, C.}, \bibinfo{author}{Vapnik, V.},
  \bibinfo{year}{1995}.
\newblock \bibinfo{title}{Support-vector networks}.
\newblock \bibinfo{journal}{Machine Learning} \bibinfo{volume}{20},
  \bibinfo{pages}{273--297}.
\newblock \URLprefix \url{https://doi.org/10.1023/A:1022627411411},
  \DOIprefix\doi{10.1023/A:1022627411411}.
%Type = Book
\bibitem[{Duda et~al.(2001)Duda, Hart and Stork}]{duda2001pattern}
\bibinfo{author}{Duda, R.O.}, \bibinfo{author}{Hart, P.E.},
  \bibinfo{author}{Stork, D.G.}, \bibinfo{year}{2001}.
\newblock \bibinfo{title}{Pattern classification}.
\newblock \bibinfo{publisher}{Wiley}, \bibinfo{address}{New York}.
%Type = Article
\bibitem[{Friedman(2000)}]{Friedman00greedyfunction}
\bibinfo{author}{Friedman, J.H.}, \bibinfo{year}{2000}.
\newblock \bibinfo{title}{Greedy function approximation: A gradient boosting
  machine}.
\newblock \bibinfo{journal}{Annals of Statistics} \bibinfo{volume}{29},
  \bibinfo{pages}{1189--1232}.
%Type = Inproceedings
\bibitem[{Ginde et~al.(2015)Ginde, Saha, Balasubramaniam, Harsha, Mathur,
  Dayasagar and Anand}]{ginde2015mining}
\bibinfo{author}{Ginde, G.}, \bibinfo{author}{Saha, S.},
  \bibinfo{author}{Balasubramaniam, C.}, \bibinfo{author}{Harsha, R.},
  \bibinfo{author}{Mathur, A.}, \bibinfo{author}{Dayasagar, B.},
  \bibinfo{author}{Anand, M.}, \bibinfo{year}{2015}.
\newblock \bibinfo{title}{Mining massive databases for computation of
  scholastic indices: Model and quantify internationality and influence
  diffusion of peer-reviewed journals}, in: \bibinfo{booktitle}{Proceedings of
  the fourth national conference of Institute of Scientometrics, SIoT}.
%Type = Article
\bibitem[{Ginde et~al.(2016)Ginde, Saha, Mathur, Venkatagiri, Vadakkepat,
  Narasimhamurthy and Daya~Sagar}]{Ginde2016}
\bibinfo{author}{Ginde, G.}, \bibinfo{author}{Saha, S.},
  \bibinfo{author}{Mathur, A.}, \bibinfo{author}{Venkatagiri, S.},
  \bibinfo{author}{Vadakkepat, S.}, \bibinfo{author}{Narasimhamurthy, A.},
  \bibinfo{author}{Daya~Sagar, B.S.}, \bibinfo{year}{2016}.
\newblock \bibinfo{title}{Scientobase: a framework and model for computing
  scholastic indicators of non-local influence of journals via native data
  acquisition algorithms}.
\newblock \bibinfo{journal}{Scientometrics} \bibinfo{volume}{108},
  \bibinfo{pages}{1479--1529}.
\newblock \URLprefix \url{https://doi.org/10.1007/s11192-016-2006-2},
  \DOIprefix\doi{10.1007/s11192-016-2006-2}.
%Type = Article
\bibitem[{Irwin et~al.(2014)Irwin, M{\'{e}}ndez, Fair{\'{e}}n and
  Schulze-Makuch}]{Irwin2014}
\bibinfo{author}{Irwin, L.}, \bibinfo{author}{M{\'{e}}ndez, A.},
  \bibinfo{author}{Fair{\'{e}}n, A.}, \bibinfo{author}{Schulze-Makuch, D.},
  \bibinfo{year}{2014}.
\newblock \bibinfo{title}{Assessing the possibility of biological complexity on
  other worlds, with an estimate of the occurrence of complex life in the milky
  way galaxy}.
\newblock \bibinfo{journal}{Challenges} \bibinfo{volume}{5},
  \bibinfo{pages}{159--174}.
\newblock \URLprefix \url{https://doi.org/10.3390/challe5010159},
  \DOIprefix\doi{10.3390/challe5010159}.
%Type = Misc
\bibitem[{Khaidem et~al.(2016)Khaidem, Saha, Basak and Dey}]{Khaidem2016}
\bibinfo{author}{Khaidem, L.}, \bibinfo{author}{Saha, S.},
  \bibinfo{author}{Basak, S.}, \bibinfo{author}{Dey, S.R.},
  \bibinfo{year}{2016}.
\newblock \bibinfo{title}{Predicting the direction of stock market prices using
  random forest}.
\newblock \URLprefix
  \url{"https://www.researchgate.net/publication/301818771_Predicting_the_direction_of_stock_market_prices_using_random_forest"}.
%Type = Article
\bibitem[{M{\'{e}}ndez(2011a)}]{hipparcosref}
\bibinfo{author}{M{\'{e}}ndez, A.}, \bibinfo{year}{2011}a.
\newblock \bibinfo{title}{The night sky of exoplanets} \URLprefix
  \url{http://phl.upr.edu/library/notes/syntheticstars}.
%Type = Article
\bibitem[{M{\'{e}}ndez(2011b)}]{thermalclassif}
\bibinfo{author}{M{\'{e}}ndez, A.}, \bibinfo{year}{2011}b.
\newblock \bibinfo{title}{A thermal planetary habitability classification for
  exoplanets} \URLprefix
  \url{http://phl.upr.edu/library/notes/athermalplanetaryhabitabilityclassificationforexoplanets}.
%Type = Article
\bibitem[{M{\'{e}}ndez(2018)}]{phlref}
\bibinfo{author}{M{\'{e}}ndez, A.}, \bibinfo{year}{2018}.
\newblock \bibinfo{title}{The habitable exoplanets catalog} \URLprefix
  \url{http://phl.upr.edu/hec}.
%Type = Article
\bibitem[{Mohanchandra et~al.(2015)Mohanchandra, Saha, Murthy and
  Lingaraju}]{Mohanchandra2015}
\bibinfo{author}{Mohanchandra, K.}, \bibinfo{author}{Saha, S.},
  \bibinfo{author}{Murthy, K.S.}, \bibinfo{author}{Lingaraju, G.},
  \bibinfo{year}{2015}.
\newblock \bibinfo{title}{Distinct adoption of k-nearest neighbour and support
  vector machine in classifying {EEG} signals of mental tasks}.
\newblock \bibinfo{journal}{International Journal of Intelligent Engineering
  Informatics} \bibinfo{volume}{3}, \bibinfo{pages}{313}.
\newblock \URLprefix \url{https://doi.org/10.1504/ijiei.2015.073064},
  \DOIprefix\doi{10.1504/ijiei.2015.073064}.
%Type = Article
\bibitem[{Parzen(1962)}]{Parzen1962}
\bibinfo{author}{Parzen, E.}, \bibinfo{year}{1962}.
\newblock \bibinfo{title}{On estimation of a probability density function and
  mode}.
\newblock \bibinfo{journal}{The Annals of Mathematical Statistics}
  \bibinfo{volume}{33}, \bibinfo{pages}{1065--1076}.
\newblock \URLprefix \url{https://doi.org/10.1214/aoms/1177704472},
  \DOIprefix\doi{10.1214/aoms/1177704472}.
%Type = Article
\bibitem[{Pedregosa et~al.(2011)Pedregosa, Varoquaux, Gramfort, Michel,
  Thirion, Grisel, Blondel, Prettenhofer, Weiss, Dubourg, Vanderplas, Passos,
  Cournapeau, Brucher, Perrot and Duchesnay}]{scikit-learn}
\bibinfo{author}{Pedregosa, F.}, \bibinfo{author}{Varoquaux, G.},
  \bibinfo{author}{Gramfort, A.}, \bibinfo{author}{Michel, V.},
  \bibinfo{author}{Thirion, B.}, \bibinfo{author}{Grisel, O.},
  \bibinfo{author}{Blondel, M.}, \bibinfo{author}{Prettenhofer, P.},
  \bibinfo{author}{Weiss, R.}, \bibinfo{author}{Dubourg, V.},
  \bibinfo{author}{Vanderplas, J.}, \bibinfo{author}{Passos, A.},
  \bibinfo{author}{Cournapeau, D.}, \bibinfo{author}{Brucher, M.},
  \bibinfo{author}{Perrot, M.}, \bibinfo{author}{Duchesnay, E.},
  \bibinfo{year}{2011}.
\newblock \bibinfo{title}{Scikit-learn: Machine learning in {P}ython}.
\newblock \bibinfo{journal}{Journal of Machine Learning Research}
  \bibinfo{volume}{12}, \bibinfo{pages}{2825--2830}.
%Type = Article
\bibitem[{Peng et~al.(2013)Peng, Zhang and Zhao}]{Peng2013}
\bibinfo{author}{Peng, N.}, \bibinfo{author}{Zhang, Y.}, \bibinfo{author}{Zhao,
  Y.}, \bibinfo{year}{2013}.
\newblock \bibinfo{title}{A {SVM}-{kNN} method for quasar-star classification}.
\newblock \bibinfo{journal}{Science China Physics, Mechanics and Astronomy}
  \bibinfo{volume}{56}, \bibinfo{pages}{1227--1234}.
\newblock \URLprefix \url{https://doi.org/10.1007/s11433-013-5083-8},
  \DOIprefix\doi{10.1007/s11433-013-5083-8}.
%Type = Article
\bibitem[{Quinlan(1986)}]{Quinlan1986}
\bibinfo{author}{Quinlan, J.R.}, \bibinfo{year}{1986}.
\newblock \bibinfo{title}{Induction of decision trees}.
\newblock \bibinfo{journal}{Machine Learning} \bibinfo{volume}{1},
  \bibinfo{pages}{81--106}.
\newblock \URLprefix \url{https://doi.org/10.1007/bf00116251},
  \DOIprefix\doi{10.1007/bf00116251}.
%Type = Inproceedings
\bibitem[{Rish(2001)}]{rish2001empirical}
\bibinfo{author}{Rish, I.}, \bibinfo{year}{2001}.
\newblock \bibinfo{title}{An empirical study of the naive bayes classifier},
  in: \bibinfo{booktitle}{IJCAI 2001 workshop on empirical methods in
  artificial intelligence}, \bibinfo{organization}{IBM New York}. pp.
  \bibinfo{pages}{41--46}.
%Type = Article
\bibitem[{Rosenblatt(1956)}]{Rosenblatt1956}
\bibinfo{author}{Rosenblatt, M.}, \bibinfo{year}{1956}.
\newblock \bibinfo{title}{Remarks on some nonparametric estimates of a density
  function}.
\newblock \bibinfo{journal}{The Annals of Mathematical Statistics}
  \bibinfo{volume}{27}, \bibinfo{pages}{832--837}.
\newblock \URLprefix \url{https://doi.org/10.1214/aoms/1177728190},
  \DOIprefix\doi{10.1214/aoms/1177728190}.
%Type = Article
\bibitem[{{Saha} et~al.(2017){Saha}, {Basak}, {Bora}, {Safonova}, {Agrawal},
  {Sarkar} and {Murthy}}]{Proxb}
\bibinfo{author}{{Saha}, S.}, \bibinfo{author}{{Basak}, S.},
  \bibinfo{author}{{Bora}, K.}, \bibinfo{author}{{Safonova}, M.},
  \bibinfo{author}{{Agrawal}, S.}, \bibinfo{author}{{Sarkar}, P.},
  \bibinfo{author}{{Murthy}, J.}, \bibinfo{year}{2017}.
\newblock \bibinfo{title}{{Theoretical Validation of Potential Habitability via
  Analytical and Boosted Tree Methods: An Optimistic Study on Recently
  Discovered Exoplanets}}.
\newblock \bibinfo{journal}{ArXiv e-prints}
  \href{http://arxiv.org/abs/1712.01040}{\tt arXiv:1712.01040}.
%Type = Article
\bibitem[{Saha et~al.(2016)Saha, Sarkar, Dwivedi, Dwivedi, Narasimhamurthy and
  Roy}]{Saha2016}
\bibinfo{author}{Saha, S.}, \bibinfo{author}{Sarkar, J.},
  \bibinfo{author}{Dwivedi, A.}, \bibinfo{author}{Dwivedi, N.},
  \bibinfo{author}{Narasimhamurthy, A.M.}, \bibinfo{author}{Roy, R.},
  \bibinfo{year}{2016}.
\newblock \bibinfo{title}{A novel revenue optimization model to address the
  operation and maintenance cost of a data center}.
\newblock \bibinfo{journal}{Journal of Cloud Computing} \bibinfo{volume}{5},
  \bibinfo{pages}{1}.
\newblock \URLprefix \url{https://doi.org/10.1186/s13677-015-0050-8},
  \DOIprefix\doi{10.1186/s13677-015-0050-8}.
%Type = Article
\bibitem[{Sale(2015)}]{Sale2015}
\bibinfo{author}{Sale, S.E.}, \bibinfo{year}{2015}.
\newblock \bibinfo{title}{Three-dimensional extinction mapping and selection
  effects}.
\newblock \bibinfo{journal}{Monthly Notices of the Royal Astronomical Society}
  \bibinfo{volume}{452}, \bibinfo{pages}{2960--2972}.
\newblock \URLprefix \url{https://doi.org/10.1093/mnras/stv1459},
  \DOIprefix\doi{10.1093/mnras/stv1459}.
%Type = Misc
\bibitem[{Sarkar et~al.(2016)Sarkar, Goswami, Saha and Kar}]{1610.00624}
\bibinfo{author}{Sarkar, J.}, \bibinfo{author}{Goswami, B.},
  \bibinfo{author}{Saha, S.}, \bibinfo{author}{Kar, S.}, \bibinfo{year}{2016}.
\newblock \bibinfo{title}{Cdsfa stochastic frontier analysis approach to
  revenue modeling in large cloud data centers}.
\newblock \href{http://arxiv.org/abs/arXiv:1610.00624}{\tt
  arXiv:arXiv:1610.00624}.
%Type = Article
\bibitem[{Schulze-Makuch and Bains(2018)}]{schulze-makuch2018time}
\bibinfo{author}{Schulze-Makuch, D.}, \bibinfo{author}{Bains, W.},
  \bibinfo{year}{2018}.
\newblock \bibinfo{title}{Time to consider search strategies for complex life
  on exoplanets}.
\newblock \bibinfo{journal}{Nature Astronomy} , \bibinfo{pages}{1--2}\URLprefix
  \url{http:https://doi.org/10.1038/s41550-018-0476-2},
  \DOIprefix\doi{10.1038/s41550-018-0476-2}.
%Type = Article
\bibitem[{Schulze-Makuch et~al.(2011)Schulze-Makuch, M{\'{e}}ndez,
  Fair{\'{e}}n, von Paris, Turse, Boyer, Davila, de~Sousa~Ant{\'{o}}nio,
  Catling and Irwin}]{SchulzeMakuch2011}
\bibinfo{author}{Schulze-Makuch, D.}, \bibinfo{author}{M{\'{e}}ndez, A.},
  \bibinfo{author}{Fair{\'{e}}n, A.G.}, \bibinfo{author}{von Paris, P.},
  \bibinfo{author}{Turse, C.}, \bibinfo{author}{Boyer, G.},
  \bibinfo{author}{Davila, A.F.}, \bibinfo{author}{de~Sousa~Ant{\'{o}}nio,
  M.R.}, \bibinfo{author}{Catling, D.}, \bibinfo{author}{Irwin, L.N.},
  \bibinfo{year}{2011}.
\newblock \bibinfo{title}{A two-tiered approach to assessing the habitability
  of exoplanets}.
\newblock \bibinfo{journal}{Astrobiology} \bibinfo{volume}{11},
  \bibinfo{pages}{1041--1052}.
\newblock \URLprefix \url{https://doi.org/10.1089/ast.2010.0592},
  \DOIprefix\doi{10.1089/ast.2010.0592}.
%Type = Article
\bibitem[{Shallue and Vanderburg(2018)}]{googlenasa}
\bibinfo{author}{Shallue, C.J.}, \bibinfo{author}{Vanderburg, A.},
  \bibinfo{year}{2018}.
\newblock \bibinfo{title}{Identifying exoplanets with deep learning: A
  five-planet resonant chain around kepler-80 and an eighth planet around
  kepler-90}.
\newblock \bibinfo{journal}{The Astronomical Journal} \bibinfo{volume}{155},
  \bibinfo{pages}{94}.
\newblock \URLprefix \url{http://stacks.iop.org/1538-3881/155/i=2/a=94}.
%Type = Article
\bibitem[{Strigari et~al.(2012)Strigari, Barnab{\`{e}}, Marshall and
  Blandford}]{Strigari2012}
\bibinfo{author}{Strigari, L.E.}, \bibinfo{author}{Barnab{\`{e}}, M.},
  \bibinfo{author}{Marshall, P.J.}, \bibinfo{author}{Blandford, R.D.},
  \bibinfo{year}{2012}.
\newblock \bibinfo{title}{Nomads of the galaxy}.
\newblock \bibinfo{journal}{Monthly Notices of the Royal Astronomical Society}
  \bibinfo{volume}{423}, \bibinfo{pages}{1856--1865}.
\newblock \URLprefix \url{https://doi.org/10.1111/j.1365-2966.2012.21009.x},
  \DOIprefix\doi{10.1111/j.1365-2966.2012.21009.x}.
%Type = Article
\bibitem[{Vapnik and Chervonenkis(1964)}]{vapnik1964}
\bibinfo{author}{Vapnik, V.N.}, \bibinfo{author}{Chervonenkis, A.Y.},
  \bibinfo{year}{1964}.
\newblock \bibinfo{title}{On a class of perceptrons}.
\newblock \bibinfo{journal}{Automation and Remote Control} \bibinfo{volume}{1},
  \bibinfo{pages}{103--109}.
%Type = Article
\bibitem[{de~Wit et~al.(2018)de~Wit, Wakeford, Lewis, Delrez, Gillon, Selsis,
  Leconte, Demory, Bolmont, Bourrier, Burgasser, Grimm, Jehin, Lederer, Owen,
  Stamenkovi{\'{c}} and Triaud}]{deWit2018}
\bibinfo{author}{de~Wit, J.}, \bibinfo{author}{Wakeford, H.R.},
  \bibinfo{author}{Lewis, N.K.}, \bibinfo{author}{Delrez, L.},
  \bibinfo{author}{Gillon, M.}, \bibinfo{author}{Selsis, F.},
  \bibinfo{author}{Leconte, J.}, \bibinfo{author}{Demory, B.O.},
  \bibinfo{author}{Bolmont, E.}, \bibinfo{author}{Bourrier, V.},
  \bibinfo{author}{Burgasser, A.J.}, \bibinfo{author}{Grimm, S.},
  \bibinfo{author}{Jehin, E.}, \bibinfo{author}{Lederer, S.M.},
  \bibinfo{author}{Owen, J.E.}, \bibinfo{author}{Stamenkovi{\'{c}}, V.},
  \bibinfo{author}{Triaud, A.H.M.J.}, \bibinfo{year}{2018}.
\newblock \bibinfo{title}{Atmospheric reconnaissance of the habitable-zone
  earth-sized planets orbiting {TRAPPIST}-1}.
\newblock \bibinfo{journal}{Nature Astronomy} \bibinfo{volume}{2},
  \bibinfo{pages}{214--219}.
\newblock \URLprefix \url{https://doi.org/10.1038/s41550-017-0374-z},
  \DOIprefix\doi{10.1038/s41550-017-0374-z}.
%Type = Inbook
\bibitem[{Zighed et~al.(2010)Zighed, Ritschard and Marcellin}]{Zighed2010}
\bibinfo{author}{Zighed, D.A.}, \bibinfo{author}{Ritschard, G.},
  \bibinfo{author}{Marcellin, S.}, \bibinfo{year}{2010}.
\newblock \bibinfo{title}{Asymmetric and Sample Size Sensitive Entropy Measures
  for Supervised Learning}. \bibinfo{publisher}{Springer Berlin Heidelberg},
  \bibinfo{address}{Berlin, Heidelberg}.
\newblock pp. \bibinfo{pages}{27--42}.
\newblock \URLprefix \url{https://doi.org/10.1007/978-3-642-05183-8_2},
  \DOIprefix\doi{10.1007/978-3-642-05183-8_2}.

\end{thebibliography}

\end{document}